\documentclass[pdflatex,sn-nature]{sn-jnl}%

\usepackage{ascmac}

\usepackage{amsmath,amssymb,amsfonts}%
\usepackage{xcolor}%
\usepackage{booktabs}%
\usepackage{algorithm}%
\usepackage[sectionbib]{bibunits}
\usepackage{bm}
\usepackage[version=3]{mhchem}
\usepackage{enumitem}
\definecolor{blue}{RGB}{25, 130, 196}

\theoremstyle{definition}
\newtheorem{define}{Definition}
\newtheorem{ex}{Example}

\begin{document}

\title{Experimental access to molarity's blind spot \\ in macroscopic assays}

\author[1]{\fnm{Fuyuki} \sur{Matsuda}}\equalcont{These authors contributed equally to this work.}
\author[1]{\fnm{Masahiko} \sur{Yoshimura}}\equalcont{These authors contributed equally to this work.}
\author[2]{\fnm{Shiro} \sur{Ikeda}}
\author*[1,3]{\fnm{Daishi} \sur{Fujita}}\email{dfujita@icems.kyoto-u.ac.jp}

\affil*[1]{\orgdiv{Institute for Integrated Cell-Material Sciences (iCeMS), Institute for Advanced Study}, \orgname{Kyoto University}, \orgaddress{\street{Yoshida, Sakyo-ku}, \city{Kyoto}, \postcode{606-8501}, \country{Japan}}}

\affil[2]{\orgname{The Institute of Statistical Mathematics}, \orgaddress{\street{10-3 Midori-cho}, \city{Tachikawa}, \state{Tokyo}, \postcode{190-8562}, \country{Japan}}}

\affil[3]{\orgname{Inamori Research Institute for Science}, \orgaddress{\city{Kyoto}, \country{Japan}}}

\abstract{Chemical kinetics has long inferred local molecular behaviour through the flask-and-molarity pairing, where well-mixed concentrations serve as the experimental readout.
Yet many biological reactions occur in structured environments.
Researchers have long recognized that concentration may not carry the same operational meaning in such environments, but even local concepts such as effective molarity usually translate local effects back into a single value with units of concentration.
What has been missing is the complementary path: a bench-compatible way to make local structure an experimental variable, rather than only a correction to molarity.
Here we show a chemistry--geometry crossover that the flask-and-molarity interface could not make visible.
In the micromolar-or-weaker affinity regime, inhibition can switch sharply out of the familiar concentration-and-affinity mode: chemical binding strength no longer determines the response, and the shape of the target's local space does.
A bench-compatible interface made this switch measurable by separating bulk dose from local geometry.
This blind spot arose from the hidden premise that macroscopic pooling makes a structured local state readable as a single local concentration.
The chemistry--geometry crossover breaks that premise: in a structured target environment, a macroscopic assay can remain sensitive to the probability distribution of local states, so collapsing that distribution to one concentration-valued number removes the geometric control axis from the readout.
By preserving that axis in the experiment, the interface bypasses molarity's hidden bottleneck and provides a routine experimental route to remeasure and reinterpret molecular interactions in structured space.}

\maketitle

The quantitative study of reaction kinetics has shaped modern molecular science for over a century\cite{Laidler1987}.
The Law of Mass Action (1860s)~\cite{Waage1864}, the Arrhenius equation (1880s)~\cite{Arrhenius1889}, and Michaelis--Menten enzyme kinetics (1913)~\cite{Michaelis1913} established the foundations of reaction kinetics.
These foundations were later given a rigorous physical basis by Gibbs\cite{Gibbs1878} and Einstein\cite{Einstein1905}, and extended by Kramers\cite{Kramers1940} (Fig.~1).
Classical reaction kinetics provides a conceptual foundation for modern biochemistry, molecular biology, and biophysics\cite{Porciani2024,Wagh2023,Nettels2024}.
Beyond these core disciplines, its central parameters ($k_{\mathrm{on}}$, $k_{\mathrm{off}}$, and $K_{\mathrm{d}}$) have become a shared vocabulary for quantifying molecular interactions across science and industry\cite{Wang2023}, particularly in drug discovery\cite{Schuetz2017,Liu2024,Pinto2024}.
The framework excels at modelling molecular interactions in bulk, homogeneous \textit{in vitro} (test-tube) systems, where concentrations are uniform and ensemble averaging yields stable, interpretable kinetics.
Yet the defining features of \textit{in vivo} environments (spatial heterogeneity\cite{Gorshkova2008,Garner2023}, low-copy-number stochasticity\cite{Bhalla2004,Ham2024}, and nanoscale confinement\cite{Erbas2019,Siddique2024}) belong to mesoscopic regimes where neither uniformity of concentration nor ensemble averaging can be assumed.
Quantifying molecular interactions in these regimes remains an open challenge.

\begin{figure}[t!]
  \begin{center}
  \includegraphics[width=1.0\textwidth]{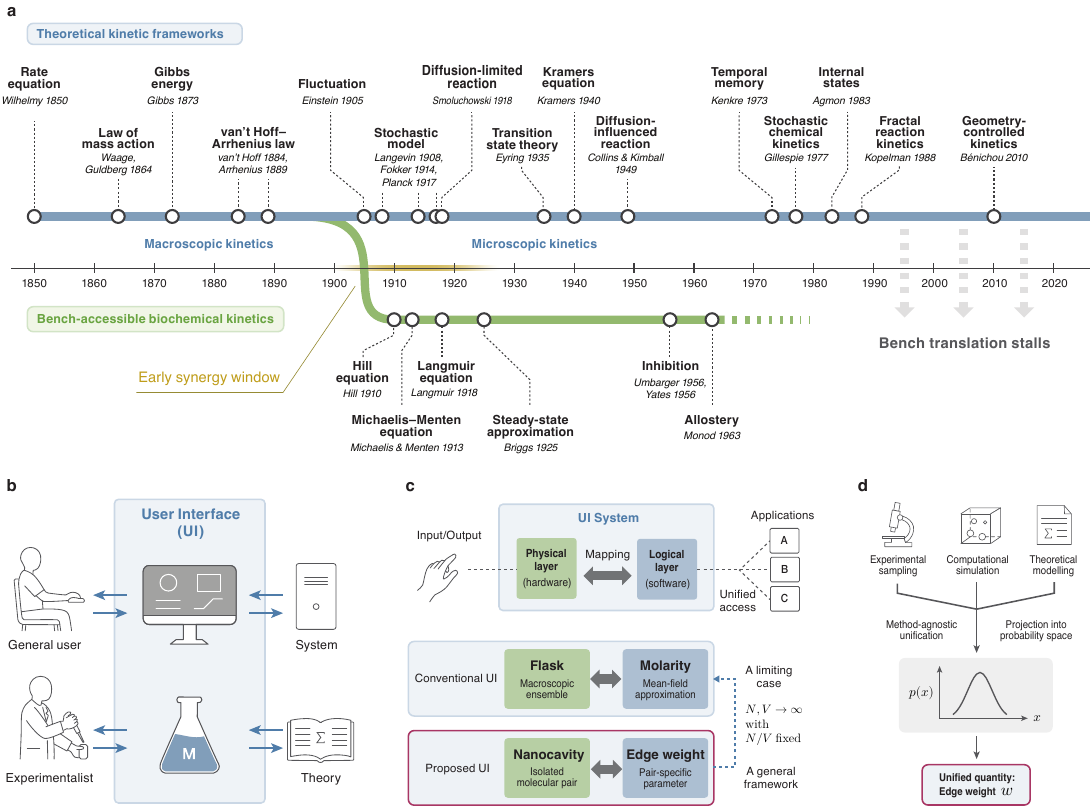}
  \caption{\textbf{A user-interface view of reaction kinetics.}
  \textbf{a}, Historical divergence between theoretical kinetic frameworks and bench-accessible biochemical kinetics.
  The upper track summarizes theoretical reaction-kinetic milestones, progressing from macroscopic rate laws to microscopic descriptions of fluctuations, diffusion, memory and state-dependent reactions\cite{Wilhelmy1850,Waage1864,Gibbs1873,vantHoff1884,Arrhenius1889,Einstein1905,Langevin1908,Fokker1914,Planck1917,Smoluchowski1918,Eyring1935,Kramers1940,Collins1949,Kenkre1973,Gillespie1977,Agmon1983,Kopelman1988,Benichou2010}.
  The lower track shows biochemical frameworks that became routine in bench practice, including Hill cooperativity, Michaelis--Menten kinetics, Langmuir binding, steady-state approximation, inhibition and allostery\cite{Hill1910,Michaelis1913,Langmuir1918,Briggs1925,Umbarger1956,Yates1956,Monod1963}.
  In the early twentieth century, physicochemical theory was translated into bench-accessible biochemical kinetics, helping establish modern quantitative biochemistry.
  Later theory expanded towards microscopic, stochastic and diffusion-influenced descriptions, but comparable theory-to-bench translation has long stalled.
  \textbf{b}, A user-interface view of kinetic experiments.
  A user interface (UI) makes a complex system usable by exposing selected inputs and outputs without requiring full knowledge of its internal mechanisms.
  By analogy, flask-based assays and molarity have served as the practical UI through which bench measurements communicate with kinetic equations.
  \textbf{c}, Physical and logical layers of kinetic access.
  The conventional kinetic UI pairs a flask-scale ensemble with molarity, a mean-field readout.
  The proposed UI pairs a nanocavity that isolates one molecular pair with an edge weight $w$ that represents its pair-specific interaction.
  The flask--molarity description is recovered in the homogeneous, large-number limit.
  \textbf{d}, Probability-density description as a common language.
  In the implementation shown here, experimental sampling, computational simulation and theoretical modelling can be mapped onto probability densities $p(x)$.
  This shared representation makes method-specific outputs mutually interpretable as descriptions of local molecular organization, unified by the edge weight $w$.}
  \label{fig1}
  \end{center}
\end{figure}

Theoretical work has directly targeted these limitations through deficiency theory\cite{Feinberg1987}, queueing approaches\cite{Zhang2019}, stochastic chemical-reaction-network formulations\cite{Anderson2010}, and spatial reaction-diffusion models\cite{Kopelman1988,Benichou2010}.
These advances, however, have yet to translate into routine experimental practice; the initial synergy between theory and experiment that built modern kinetics has not been sustained (Fig.~1).
This disconnect arises not primarily from deficiencies in the theories themselves, but from a structural mismatch between what theory can describe and what experiment can access.
The parameters that modern frameworks require are typically difficult to measure and even harder to control.
Elegant though they are, these theories therefore offer experimentalists no routine, general-purpose path; the bottleneck is an interface problem.

This interface problem finds a direct parallel in the concept of the User Interface (UI) from Human-Computer Interaction\cite{ISO9241-110-2020}.
A UI governs which capabilities users can access, without requiring mastery of the system's internal complexity.
In kinetics, the classic flask-and-molarity pairing has long served as the \textit{de facto} UI between theory and experiment.
Yet this very familiarity masks a bottleneck: because the flask-and-molarity interface presupposes spatial uniformity, it does not support modern theories designed for heterogeneous, nanoscale environments.
Thus, a UI functions as a scientific constraint on what is natural to specify and measure; it shapes the boundaries of exploration, and apparent experimental limitations often reflect limitations of the UI itself.

We therefore treat flask-based molarity as one scientific interface rather than the natural endpoint of kinetic description, and ask what interface is needed for micro- and mesoscopic reaction regimes.
The framework proposed here is a meta-level wrapper, not a competing kinetic theory: it recasts parameters from existing formalisms as experimentally adjustable edge weights in a weighted graph.
In the logical layer, entities are vertices and pairwise encounters are edges; in the physical layer, reconfigurable DNA nanostructures implement those edges as molecular-pair readouts.
The design follows the user-centred logic of interface design\cite{ISO9241-210-2019}: the relevant variables must remain measurable, tunable and interpretable at the bench.
We use this interface to make local structure an experimental variable and to probe a blind spot of molarity in macroscopic assays.
A companion study translates the same principle into practically important molecular-interaction settings, offering a concrete implementation that may also serve as an intuitive entry point to the framework developed here.

\section*{Graph-theoretic common language for kinetic theories}

The theoretical layer of this scientific UI must accommodate kinetic models that currently reside in mutually isolated frameworks.
We introduce graph-theoretic notation at the level of individual interacting entities (such as ions, molecules, or larger assemblies): each entity is represented as a vertex, and each candidate pairwise encounter as an edge.
By recasting reactions and interactions as collections of candidate pairwise encounters, this graph lets different kinetic formalisms express their contributions in comparable terms.
By accommodating each theory's distinct edge definitions within one structure, the graph places their rate laws on common ground, ready for formal and experimental comparison.

Even the classical bimolecular rate law encodes a latent graph structure (Fig.~2a), one that this pairwise perspective makes explicit.
Consider a familiar binary association, $X_1 + X_2 \rightleftarrows X_3$: mass action gives $r_{X3} = k_c[X_1][X_2]$, where $k_c$ is the concentration-based rate constant and $[X_i]$ denotes the molar concentration of entity class $X_i$ (such as a chemical species).
In the well-mixed limit implicit in this rate law, each of the $N_{X1}$ entities in class $X_1$ can potentially encounter all $N_{X2}$ entities in class $X_2$, yielding $N_{X2}$ candidate partners per $X_1$ entity and $N_{X1} \cdot N_{X2}$ candidate pairs in total.
These candidate pairs map to edges of a complete bipartite graph, with one node set for each entity class and every cross-class pair connected, giving $|E| = N_{X1} \cdot N_{X2}$.
Since molarity relates to entity count via $[X_i] = N_{X_i} / (V N_{A})$, with $V$ the system volume and $N_A$ Avogadro's number, the rate law can be rewritten as:
\begin{equation}
r_{X3} = k_c[X_1][X_2] = k |E|
\label{eq_classical_rate_law}
\end{equation}
where $k = k_c/(V^{2}N_A^{2})$ absorbs the concentration-to-count conversion.
The classical rate is thus exactly proportional to the number of graph edges; concentration notation packages what is, at root, a discrete count of candidate interaction pairs.
This equivalence holds under one implicit assumption: every pair contributes equally to the overall rate, a limitation relaxed next by allowing heterogeneous edge weights (Fig.~2b).

\begin{figure}[t!]
  \begin{center}
  \includegraphics[width=1.0\textwidth]{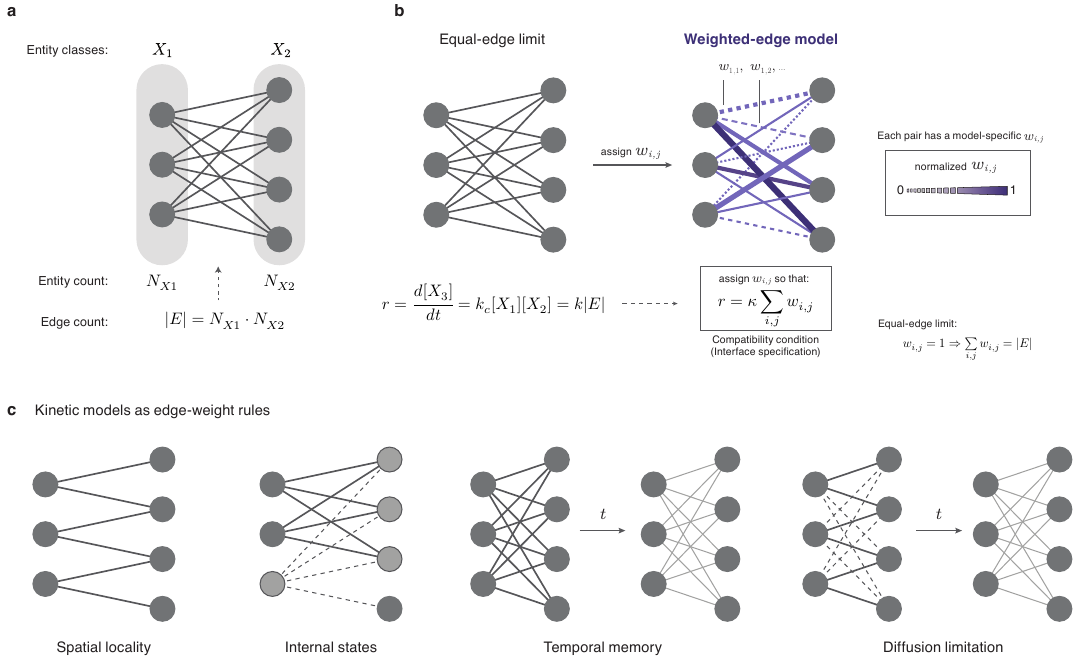}
  \caption{\textbf{A weighted-edge framework for reaction kinetics.}
  \textbf{a}, Binary association as a graph.
  For the binary association $X_1 + X_2 \rightleftarrows X_3$, an entity is an individual interacting object, such as an ion, molecule, or larger assembly, and an entity class is a model-defined grouping of such objects, based for example on chemical identity, internal state, or spatial localization.
  Each entity is a node, and each possible $X_1$--$X_2$ pairing is an edge.
  In the well-mixed limit, every $X_1$ entity can pair with every $X_2$ entity, giving the edge count $|E| = N_{\text{X1}} \cdot N_{\text{X2}}$.
  \textbf{b}, From concentration product to weighted edges.
  In the equal-edge limit, all possible $X_1$--$X_2$ pairings contribute identically.
  After concentration-to-count conversion, the mass-action term $k_c[\text{X}_1][\text{X}_2]$ becomes $k|E|$, where $k$ absorbs the volume and Avogadro-number factors; the concentration product is therefore proportional to the number of possible $X_1$--$X_2$ edges in the well-mixed limit.
  For heterogeneous models, $w_{i,j}$ is assigned so that the compatibility condition $r = \kappa \sum_{i,j} w_{i,j}$ holds, where $\kappa$ is the global scale of the weighted-edge model.
  This condition is the interface specification: the additive graph form remains fixed, while each model enters through its own rule for assigning rate-compatible weights.
  \textbf{c}, Kinetic models as edge-weight rules.
  Spatial locality\cite{Bhalla2004,Peletier2003}, internal states\cite{McCammon1981,Zhou1996}, temporal memory\cite{Montrol1965,Kenkre1973}, and diffusion limitation\cite{Smoluchowski1918,Collins1949} illustrate how distinct kinetic mechanisms can be expressed as different rules for assigning $w_{i,j}$.
  The weighted-edge framework places these otherwise separate kinetic descriptions on common mathematical ground.}
  \label{fig2}
  \end{center}
\end{figure}

We formalize this relaxation as a minimal interface specification for the framework's theoretical layer (Fig.~2b), prescribing only what models must satisfy, not how.
A kinetic model is compatible with this interface if it can express the reaction rate in the additive form:
\begin{equation}
r = \kappa \sum_{i,j} w_{i,j}
\label{eq_additivity_axiom}
\end{equation}
where $w_{i,j}$ represents the interaction weight between entities $i$ and $j$.
By convention, $w_{i,j}$ is dimensionless and encodes only pairwise interaction structure; all system-wide kinetic prefactors are absorbed into the global scale $\kappa$.
In the unweighted limit ($w_{i,j} \equiv 1$), Eq.~\eqref{eq_additivity_axiom} reduces to $r = \kappa|E|$, recovering the functional form of Eq.~\eqref{eq_classical_rate_law} up to a prefactor that depends on the encounter model.
This interface generalizes Eq.~\eqref{eq_classical_rate_law}: it admits any model whose rate decomposes into pairwise contributions, imposing no constraint on how each $w_{i,j}$ is defined.
The additive form supports modular analysis: for independent subsystems (sharing no edges), contributions sum exactly; for overlapping subsystems, global behaviour can be reconstructed once the aggregation rule is specified.
The interface accommodates kinetic models encoding spatial locality, internal states (e.g., conformational changes), temporal memory (history-dependent rates), and diffusion limitation (Fig.~2c), each expressible as a specific definition of $w_{i,j}$.
This simple axiom extends beyond pairwise graphs to temporal\cite{HolmeSaramaki2013}, multilayer\cite{ArtimeEtAl2022}, and higher-order network structures\cite{Battiston2021}.
By separating global aggregation ($\kappa$) from local physics ($w_{i,j}$), Eq.~\eqref{eq_additivity_axiom} offers a common mathematical language in which different models interoperate while retaining their mechanistic specificity, reducing model specification to a single task: defining $w_{i,j}$.

To give this abstract framework concrete form, we adopt a natural working model for $w_{i,j}$ that illustrates its predictive reach: a dimensionless encounter probability $p_{\mathrm{enc}}$ derived from the spatial overlap of entity distributions.
For entities differing only in position, this overlap is purely geometric:
\begin{equation}
w_{i,j} := p_{\mathrm{enc}} = \int d^3\mathbf{x}_1\, d^3\mathbf{x}_2\, f_{1,i}(\mathbf{x}_1;t)\, f_{2,j}(\mathbf{x}_2;t)\, \mathrm{Enc}(\mathrm{encounter};\, \mathbf{x}_1, \mathbf{x}_2, t)
\label{eq_weight}
\end{equation}
Here $f_{1,i}$ and $f_{2,j}$ are spatial probability distributions for entities $i$ and $j$, and $\mathrm{Enc}$ is the encounter criterion, generalizing the point-particle collision integral to finite-size molecular encounters (Supplementary Section~\ref{ap_Enc_extention}).
In classical collision theory, rates factor into collision frequency times reaction probability; the overlap integral provides the collision-frequency analog.
For simplicity, this working model assumes statistical independence via the factored form $f_{1,i} \cdot f_{2,j}$, though a joint distribution $f_{(1,i),(2,j)}$ applies equally well for correlated systems.

Like molarity, which depends on entity count and volume rather than chemical identity, these overlap-based weights are identity-agnostic, determined entirely by spatial distributions.
This sole dependence on spatial distributions makes the framework broadly inclusive: particle-tracking microscopy data, molecular-dynamics trajectories, and analytical models all yield the spatial distributions that enter Eq.~\eqref{eq_weight} (Fig.~1d).
In the pedagogically useful limit of uniform distributions, the integral reduces to geometric overlap, $w \propto V_{\text{X1} \cap \text{X2}}/(V_{\text{X1}}\,V_{\text{X2}})$ ($\propto$: proportional to), making the abstract weight tangible (Fig.~3a,~3b).
Although this paper focuses on spatial overlap, the additive interface admits alternative weight definitions, from contact times and energy landscapes to data-driven descriptors.

\section*{Physical implementation through engineered nanostructures}

To interface our theoretical abstraction layer with the physical world, we co-designed an experimental platform that isolates and controls single molecular pairs, each representing an individual graph edge (Fig.~3a, left).
The Zonal Engineered Nano (ZEN) system realizes this mapping: each ZEN cavity confines exactly one target molecular pair, physically isolating it from neighbouring pairs while permitting binding-state measurement (Fig.~3a).
The confinement is through-hole rather than sealed, keeping the readout site continuous with the surrounding solution.
Built using DNA nanotechnology, a mature and widely accessible technology even available through commercial custom synthesis services~\cite{Dunn2020}, the platform permits precise, programmable positioning of components with sub-nanometer design accuracy (Fig.~3a, right).
The molecules were tethered using monodisperse linkers from precision organic synthesis, avoiding the length heterogeneity of typical polymers.
Reaction sites and linker attachments occupy opposite termini in an antipodal arrangement, letting each molecule explore the full range defined by its linker length without compromising reactivity (details in~\cite{CompanionPaper}).
This exceptional control over each molecule's accessible volume and relative placement enables systematic variation of the spatial parameters that govern each edge weight.
Through this platform, the abstract edge weight $w$ becomes a tunable, measurable quantity, connecting theoretical prediction to experiment.

\begin{figure}[t!]
  \begin{center}
  \includegraphics[width=1.0\textwidth]{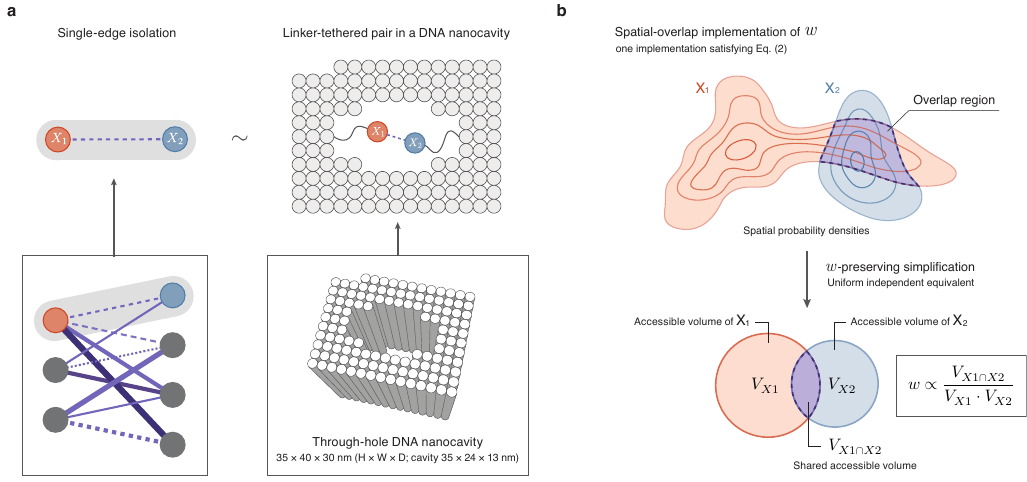}
  \caption{\textbf{Physical implementation of a tunable single-edge weight.}
  \textbf{a}, Single-edge isolation in a DNA nanocavity.
  A graph edge (left) is implemented as one linker-tethered pair, $X_1$ and $X_2$, confined in a through-hole DNA nanocavity (right).
  White circles denote DNA-duplex cross-sections forming the cavity wall.
  The cavity remains continuous with the surrounding solution but separates this pair from neighbouring pairs, removing cross-edge terms in the readout.
  Linker length, attachment geometry and accessible volume tune the corresponding edge weight $w$.
  \textbf{b}, Spatial-overlap implementation of $w$.
  Equation~\eqref{eq_additivity_axiom} defines the general additive compatibility condition; the spatial-overlap construction shown here is one implementation that satisfies it.
  In this implementation, $w$ is computed from the overlap between the spatial probability densities of $X_1$ and $X_2$ (Eq.~\eqref{eq_weight}).
  In the uniform-independent limit, this overlap reduces to $w \propto V_{\text{X1} \cap \text{X2}}/(V_{\text{X1}} \cdot V_{\text{X2}})$: the shared accessible volume normalized by the accessible volumes of $X_1$ and $X_2$.
  Distributions or geometries that preserve this normalized overlap are therefore $w$-equivalent.}
  \label{fig3}
  \end{center}
\end{figure}

To validate $w$ experimentally, we compared predicted and measured interaction patterns in ZEN cavities, using a split-luciferase reporter that signals complex formation through activity restoration (Fig.~4a; for the detailed assay configuration, see the companion paper\cite{CompanionPaper}, Fig.~4).
We computed $w$ from spatial overlap using a deliberately minimal Monte Carlo simulation that modeled molecules as tethered spheres with antipodal point distributions (Fig.~4b; Supplementary Information).
Despite omitting molecular shape, linker elasticity, and electrostatics, predicted and observed normalized signals matched closely (Pearson $r = 0.994$ on natural-log-transformed values, $n = 10$ conditions; Fig.~4c).
The root-mean-square error of $\ln(\mathrm{observed}/\mathrm{predicted})$ was $0.087$, corresponding to a typical $1.09$-fold deviation.
That such a simplified model captures experimental behaviour across linker-length combinations confirms $w$ as an effective interface between theory and experiment.

\begin{figure}[t!]
  \begin{center}
  \includegraphics[width=0.5\textwidth]{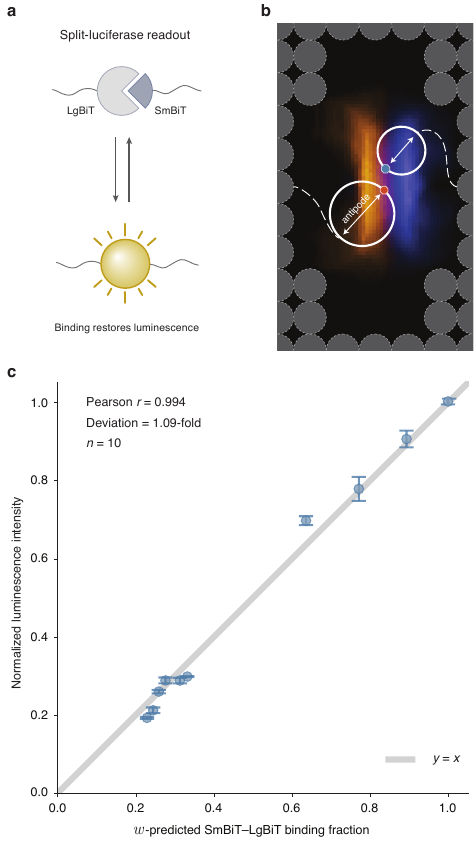}
  \caption{\textbf{Experimental validation of a minimal edge-weight model.}
  \textbf{a}, Split-luciferase complementation reporter.
  SmBiT and LgBiT are inactive when separated, but reconstitute luciferase activity upon binding, producing luminescence as a readout of complex formation.
  \textbf{b}, Minimal Monte Carlo model in a DNA nanocavity.
  The tethered proteins are approximated as spheres, with dashed curves indicating linkers.
  Coloured points mark the antipodal reporter sites, opposite the tether attachment points.
  Heatmaps show the simulated spatial distributions of these reporter sites inside the nanocavity; their overlap defines the edge weight $w$, from which the SmBiT--LgBiT binding fraction is predicted.
  \textbf{c}, $w$-based predictions versus measured luminescence across tethering geometries. Each plotted condition is one ZEN-cavity geometry defined by a specific linker-length set for SmBiT, LgBiT and the non-luminescent competitor component, with the detailed configurations reported in the companion paper\cite{CompanionPaper}. The x-axis shows the $w$-based predicted SmBiT--LgBiT binding fraction for that geometry, and the y-axis shows the normalized luminescence measured from the same construct. Points show mean $\pm$ s.d. from three replicates; $n=10$ plotted conditions after excluding the normalization reference. Despite omitting molecular shape, linker elasticity and electrostatic interactions, the predictions match the observed normalized signals closely (Pearson $r=0.994$; $\mathrm{RMSE}_{\ln}=0.087$, corresponding to a 1.09-fold typical deviation; statistics calculated on natural-log-transformed values).}
  \label{fig4}
  \end{center}
\end{figure}

These results establish the edge weight $w$ as a quantitative descriptor of pairwise reactivity, predictable from molecular geometry and tunable independently of bulk concentration.
The companion paper\cite{CompanionPaper} exploits this quantitative control to detect protein--protein interactions with $K_{\mathrm{d}}\approx10^{-2}~\mathrm{M}$ ($1~\mathrm{M}=1~\mathrm{mol/L}$), more than an order of magnitude beyond conventional detection limits, without requiring impractically high solute concentrations.
Bulk concentration cannot reach this regime; geometric control of $w$ can.
The edge weight captures local kinetic effects that bulk concentration models average out, making local structure experimentally addressable rather than hidden inside a bulk-dose readout.

\section*{Chemistry--Geometry Crossover in Structured-Space Inhibition}

The redesigned interface separates geometry from bulk dose and exposes a choice that conventional concentration-based assays hide.
The familiar path is a concentration-first view: local geometry may affect a reaction, but its effect is folded back into a scalar concentration-like quantity, as in effective molarity or effective concentration, where a local process is expressed relative to an analogous intermolecular one\cite{IUPAC_effective_molarity}.
Under that view, geometry remains within the concentration frame: it may change the scale of a dose response, but the response is still organized along a concentration axis.
The experiment below tests whether that familiar path is enough.
Either geometry remains a correction within the concentration frame, or it steps outside that frame and begins to organize inhibition on its own terms.

To make this choice experimentally concrete, we used DarkBiT inhibition of an LgBiT/SmBiT split-luciferase target pair.
DarkBiT is a catalytically silent SmBiT analogue generated by replacing an active-site residue, so it competes at the LgBiT interface without producing luminescence.
The pair was placed in a through-hole DNA nanostructure, where DNA columns laterally frame the readout site while the pore remains continuous with the bulk reservoir.
Within this bulk-connected geometry, we varied target-pair separation to tune $w$, and used SmBiT variants spanning orders of magnitude in $K_{\mathrm{d}}$ to tune the chemical scale of DarkBiT competition.
In a conventional inhibitor titration, these perturbations would be read as horizontal shifts in apparent potency along the inhibitor-dose axis, as in the original LgBiT/SmBiT NanoBiT readout\cite{Dixon2016_NanoBiT}.

The data take the second path, and the crossover is visible before any equation is invoked.
In the strong-affinity branch, the apparent IC$_{50}$ follows the familiar horizontal branch of inhibitor titration.
As SmBiT binding weakens beyond the micromolar scale ($K_{\mathrm d}\gtrsim 10^{-6}\,\mathrm{M};\,\gtrsim 1\,\mu\mathrm{M}$), however, the trajectory makes a near-right-angle turn and collapses into a vertical branch (Fig.~5).
Along this branch, neither SmBiT affinity nor target-pair separation continues to rank the apparent inhibition point on a single concentration axis.
The organizing variable has changed.
Inhibition crosses from a chemistry-ordered titration to a geometry-organized response.
Geometry is no longer folded into an effective-concentration correction; it steps outside the concentration frame and organizes inhibition on its own terms.

The vertical branch puts the usual local-concentration idea under a direct test: can the DarkBiT molecules within interaction range be treated as one local concentration, or can no single number represent them?
Let $\boldsymbol{N}_{\mathrm D}$ denote this fluctuating molecule count; throughout this section, $\langle\cdot\rangle$ denotes averaging over its distribution.
Let $W_{\mathrm{free}}$ and $W_{\mathrm{SmBiT}}$ be unnormalized local-state weights for the free and SmBiT-bound luminescent states, and let $W_{\mathrm{DarkBiT}}^{(1)}$ be the per-DarkBiT competitive weight, all constructed from $w$ and binding parameters.
If such a one-number local concentration exists, replacing $\boldsymbol{N}_{\mathrm D}$ by its mean gives the SmBiT-bound probability:
\begin{equation}
P_{\mathrm{bound}}^{(\mathrm{scalar})}
  = \frac{W_{\mathrm{SmBiT}}}
         {W_{\mathrm{free}}+W_{\mathrm{SmBiT}}
          +\langle \boldsymbol{N}_{\mathrm D}\rangle\, W_{\mathrm{DarkBiT}}^{(1)}}.
\label{eq_p_scalar}
\end{equation}
If no single local number exists, the DarkBiT count must be kept as a distribution.
The local SmBiT-bound probability is then computed for each possible count and averaged as the final step:
\begin{equation}
P_{\mathrm{bound}}^{(\mathrm{dist})}
  = \left\langle
      \frac{W_{\mathrm{SmBiT}}}
           {W_{\mathrm{free}}+W_{\mathrm{SmBiT}}
            +\boldsymbol{N}_{\mathrm D}\, W_{\mathrm{DarkBiT}}^{(1)}}
    \right\rangle.
\label{eq_p_dist}
\end{equation}
Both expressions compute the same SmBiT-bound probability from the same local-state weights.
The only difference is where the average is taken.
The data select the second route: averaging first gives only a rational concentration response, whereas averaging last preserves the distribution and leads to the hypergeometric response that captures the near-right-angle turn from horizontal titration to vertical crossover.

The hypergeometric response first appears under a natural Poisson-count assumption.
When the DarkBiT count within interaction range is small and Poisson-like, the final average in $P_{\mathrm{bound}}^{(\mathrm{dist})}$ becomes a confluent hypergeometric response, written as ${}_1F_1$ (Supplementary Section~\ref{ap_inhibition_characteristics}).
This ${}_1F_1$ response follows the horizontal branch in the strong-affinity regime, turns sharply at the micromolar-scale boundary, and drops into the vertical branch in Fig.~5.
The predicted ${}_1F_1$ landscape quantifies this collapse: within the vertical branch, even a $\sim 10^4$-fold change in $K_{\mathrm d}$ can shift the apparent inhibition point by only $\sim 1\%$.
A rational scalar expression does not reproduce this near-right-angle turn and drop.
The ${}_1F_1$ form therefore marks the Poisson-count regime of a broader distributional description; changing the count law changes the hypergeometric branch, not the need to retain the distribution.

Actual nanostructures add a constraint that the Poisson-count case leaves out: finite local capacity.
The ${}_1F_1$ case isolates the clean horizontal-to-vertical crossover, but a readout site cannot accommodate unlimited DarkBiT-accessible states.
For high-affinity SmBiT variants ($K_{\mathrm d}$ in the nanomolar-to-subnanomolar range), inhibition requires a larger local DarkBiT load.
In the high-loading regime, finite capacity changes the count law from Poisson-like to binomial-like, and the same final-averaging calculation gives a Gauss hypergeometric response, written as ${}_2F_1$ (Supplementary Section~\ref{ap_inhibition_characteristics}).
In Fig.~5, the resulting ${}_2F_1$ response preserves the crossover while adding the finite-capacity signature: the apparent-IC$_{50}$ markers bend upward after the vertical segment, and high DarkBiT doses leave residual signal rather than complete inhibition.
Thus, ${}_1F_1$ and ${}_2F_1$ are not competing fits; they are count-law specializations of the same distributional description, with ${}_2F_1$ carrying the clean crossover into the finite-capacity regime.

The redesigned interface shows that the familiar concentration response is a projection of a richer distributional picture, not a single regime.
Homogeneous bulk is the special case in which that picture collapses to one scalar value.
Outside that uniform limit, the local interaction state may have to be treated as a probability distribution, and its effect can persist even after macroscopic pooling.
The chemistry--geometry crossover is one case in which this hidden distribution becomes experimentally visible and begins to govern molecular interactions.
Concentration-like scalars remain useful as limits of the distributional picture, but they should not serve as the default coordinate in structured space.

\begin{center}
\includegraphics[width=1.0\textwidth]{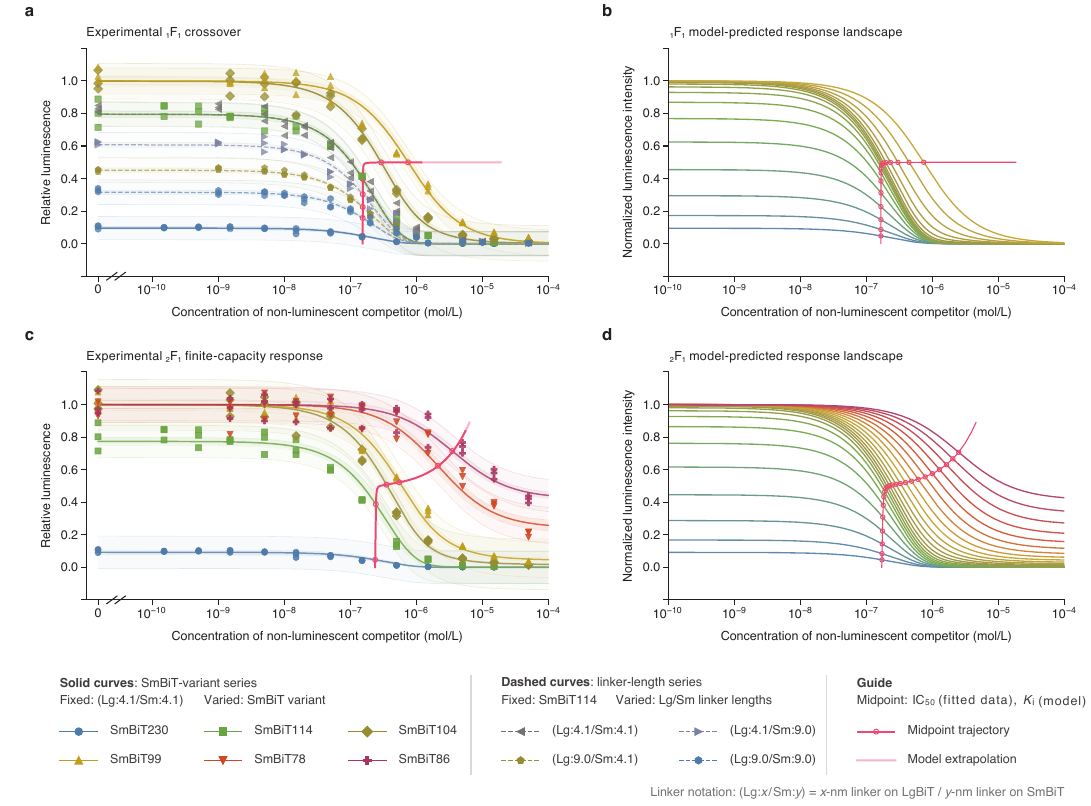}
\end{center}
\makeatletter\def\@captype{figure}\makeatother
  \caption{\textbf{Experimental chemistry-geometry crossover in structured-space inhibition.}
  In conventional inhibition, changes in binding chemistry shift response midpoints mainly along the inhibitor concentration axis.
  Here, fitted midpoints instead form a path that turns from a horizontal, chemistry-ordered branch into a near-vertical, geometry-organized branch, revealing a chemistry-geometry crossover that concentration-affinity ordering alone cannot capture.
  Open pink circles mark midpoints, defined as $\mathrm{IC}_{50}$ values in fitted experimental curves and $K_i$ values in predicted response landscapes.
  Pink paths connect these midpoints, and pale pink segments denote model extrapolation.
  \textbf{a,b}, ${}_1F_1$ Poisson-count regime.
  \textbf{a}, Experimental DarkBiT titrations of DNA-nanocavity-confined LgBiT/SmBiT split-luciferase target pairs fitted with the ${}_1F_1$ response.
  DarkBiT is a non-luminescent competitor for LgBiT; luminescence reports the fraction of LgBiT paired with SmBiT, whereas LgBiT paired with DarkBiT or left unbound is non-luminescent.
  Markers show relative luminescence measurements, and coloured curves show fitted responses.
  Solid curves vary the SmBiT variant with LgBiT and SmBiT linkers fixed at 4.1 nm; dashed curves keep SmBiT114 fixed and vary the LgBiT/SmBiT linker-length combination, as encoded in the bottom legend.
  SmBiT variants primarily change binding chemistry through $K_d$, whereas linker lengths primarily change geometry through $w$; the model combines these inputs into a common dimensionless control ratio, allowing both series to occupy the same response landscape.
  The fitted midpoint path shows the crossover directly in the experimental data.
  \textbf{b}, The corresponding ${}_1F_1$ predicted response landscape removes experimental scatter and, in the Poisson-like local competitor-count regime, reproduces the horizontal-to-vertical trajectory in \textbf{a}. In this predicted response, parts of the vertical branch are nearly affinity-invariant: $K_{\mathrm d}$ varies by $\sim 10^4$-fold, whereas the apparent $K_i$ changes by only $\sim 1\%$.
  \textbf{c,d}, ${}_2F_1$ finite-capacity regime.
  \textbf{c}, Experimental titrations fitted with the ${}_2F_1$ response using the expanded SmBiT series, including higher-affinity variants that require larger local DarkBiT loading.
  Signal scaling and condition encoding are as in \textbf{a}.
  The chemistry-geometry crossover persists, and high-loading conditions leave residual luminescence at high DarkBiT concentration rather than complete inhibition.
  \textbf{d}, The corresponding ${}_2F_1$ predicted response landscape incorporates finite local competitor capacity into the same distributional inhibition model.
  It preserves the crossover and predicts the upward bend after the vertical segment, matching the finite-capacity behaviour in \textbf{c}.
  For \textbf{a} and \textbf{c}, luminescence signals were scaled by the intrinsic activity of the matched intact, non-split luciferase construct for each SmBiT variant and then expressed relative to the fitted response scale of the corresponding titration curve.
  This normalization removes variant-specific reporter brightness while preserving vertical differences that reflect how strongly each construct reconstitutes the target complex.
  Dark shaded bands denote 95\% credible intervals for the fitted mean response, whereas pale shaded bands denote 95\% posterior predictive intervals for individual observations.}
  \label{fig5}

\section*{Connection to Established Kinetic Theories}

Scalar concentration-like variables re-emerge as derived limits when local heterogeneity is averaged out.
In the bulk limit of uniform distributions and large entity counts, the weighted graph framework (WGF) recovers the law of mass action as one such limit (Supplementary Section \ref{SI_wij}).
Does scalar recovery require this conventional bulk regime?

The Michaelis--Menten form emerges even at the single-molecule limit.
Within the ZEN cavity confining a single molecular pair, the encounter weight replaces entity count as the primitive variable.
Just as molarity divides entity count by system volume, the effective concentration divides encounter weight by interaction volume:
$$
C_{\text{eff}}=\frac{w}{\xi_{\text{enc}}},
$$
where $\xi_{\text{enc}}$ is the interaction volume obtained by integrating the encounter function.
For comparison with conventional dissociation constants ($K_{\mathrm{d}}$), defining the molar effective concentration $c_{\text{eff}} := C_{\text{eff}}/N_{\mathrm{A}}$ ($N_{\mathrm{A}}$: Avogadro's constant) yields the binding probability in the Michaelis--Menten form:
\begin{equation}
P_{\text{bound}}=\frac{c_{\text{eff}}}{K_{\mathrm{d}}+c_{\text{eff}}}.
\label{eq_michaelis_menten_form}
\end{equation}
This expression maps to the normalized Michaelis--Menten equation $v/V_{\max}=[S]/(K_M+[S])$, with $c_{\text{eff}}$ playing the role of $[S]$, $K_{\mathrm{d}}$ that of $K_M$, and $P_{\text{bound}}$ that of $v/V_{\max}$.
The standard derivation of this saturation law requires steady-state and substrate-excess conditions; here the same functional form follows from equilibrium alone, without ensemble averaging.
Just as bulk averaging recovers a scalar concentration in the macroscopic limit, pairwise isolation within the redesigned interface recovers one at the microscopic extreme.
Unlike traditional molarity, $C_{\text{eff}}$ is calculable directly from spatial distributions of the interacting entities, giving the long-sought ``local concentration'' \cite{Krishnamurthy2007} a first-principles scalar definition at the single-pair limit and connecting that microscopic scalar to the familiar bulk one.

The classical molarity-based rate law, viewed through this same framework, reveals a hidden aspect of its construction.
If Eq.~\eqref{eq_additivity_axiom} is taken as the kinetic starting point for the bimolecular reaction $X_1+X_2\rightarrow X_3$, the rate arises from the sum over pairwise encounter weights.
In a homogeneous, well-mixed system sharing a volume $V$, all pairs are statistically equivalent, and Eq.~\eqref{eq_weight} with uniform independent distributions ($f=1/V$) gives $w_{ij}=\xi_{\mathrm{enc}}/V$ (Supplementary Section~\ref{subsec:rate_equation_normalization}).
Absorbing $\xi_{\mathrm{enc}}$ into the prefactor yields
\begin{equation}
\frac{d\langle N_{X3} \rangle}{dt} = \kappa \langle N_{X1} \rangle \langle N_{X2} \rangle \frac{1}{V}
\label{eq_rate_particle_counts}
\end{equation}
in the well-mixed, large-number limit.
The familiar molarity-based equation emerges only after dividing by $V$ once more:
\begin{equation}
\frac{d[X_3]}{dt} = k_c [X_1][X_2].
\label{eq_rate_molarity_based}
\end{equation}
This closure to a single scalar requires two specific conditions: the encounter weight must already contain a factor of $1/V$, and all interacting entities must share the same homogeneous volume.
When $\xi_{\mathrm{enc}}$ becomes strongly position-dependent, or when interacting entities no longer share a common volume, $w$ does not reduce to a simple $\xi_{\mathrm{enc}}/V$ and the scalar closure breaks.
The point is easy to miss because the closure works so well in homogeneous bulk: the first $1/V$ is already present in the encounter weight, but bulk homogeneity renders it invisible.
Molarity's apparent primitiveness therefore reflects how thoroughly flask conditions satisfy these requirements, a success whose scope does not extend to the structured and nanoscale environments that modern biology investigates.
Both bulk molarity and single-pair $C_{\mathrm{eff}}$ arise as scalar closures of this pairwise structure, each under its own regime-specific conditions.

The encounter weight $w$ also carries thermodynamic meaning.
Consider the case of an equilibrium system with independent, uniformly distributed entities within their accessible volumes in the ZEN system (Fig.~3b).
Here, the positional entropy change upon binding is
\begin{equation}
\Delta S = k_{\mathrm{B}} \log w
\label{eq_entropy_change}
\end{equation}
where $k_{\mathrm{B}}$ is the Boltzmann constant.
Experimentally tuning $w$, for example by varying linker length or attachment position, directly adjusts the entropic driving force of the reaction, a degree of control that conventional bulk experiments do not provide.
For more general distributions the expression acquires additional entropy terms, but remains explicitly computable within the framework and consistent with standard statistical-mechanical treatments, including state counting, differential entropy under local equilibrium, and free-energy minimization (Supplementary Information).
Across encounter kinetics, entropic driving forces, and these distinct formalisms, $w$ recovers established results in each case.
Such consistency across independent domains suggests that $w$ serves not as a convenient parameterization of a single experimental system, but as a versatile foundation for quantifying molecular interactions.

\section*{Summary and Future Perspectives}

The flask-and-molarity UI made reaction kinetics powerful because, within its native regime, an entire reaction system could be closed onto one scalar coordinate: concentration.
The chemistry--geometry crossover reveals the regime that this successful closure had kept out of view.
In structured target environments, even a macroscopic pooled assay can cross into a branch where the local geometry of the target, rather than molecular affinity, dictates inhibition.
The blind spot is the assumption behind that reading: however structured the local states may be, macroscopic pooling should make them readable as one concentration-valued number.
Its overextension was a molarity-centred belief: local environment could be treated as no more than a correction to concentration, and stochasticity as a microscopic concern irrelevant to the macroscopic assay.
That belief was plausible not because scalar closure is fundamental, but because flask conditions make one hidden dimensional coincidence exact: a common accessible volume normalises both entity counts and encounter statistics.

When local structure breaks the flask coincidence, scalar closure is no longer guaranteed even through pooling.
Just as quantum mechanics led scientists to recognize that supposedly definite values are actually probability distributions, local concentration in structured space should now be understood in the same terms.
At the microscopic scale, distributional treatment is natural; the surprise is not that distributions exist, but that a pooled macroscopic assay can still read them.
Yet for bench scientists trained on scalar concentration, working directly with probability distributions is unintuitive.
We therefore developed the ZEN system as a practical counterpoint to the flask: by engineering the single-pair limit, ZEN turns the same distributional picture back into a scalar computable from first principles, $C_{\text{eff}}$, linking microscopic encounters to familiar bulk molarity.
In this view, the distribution is primary; familiar molarity and ZEN-derived $C_{\text{eff}}$ are scalar closures reached from opposite limits.

\begin{figure}[t!]
  \begin{center}
  \includegraphics[width=\textwidth, height=0.64\textheight, keepaspectratio]{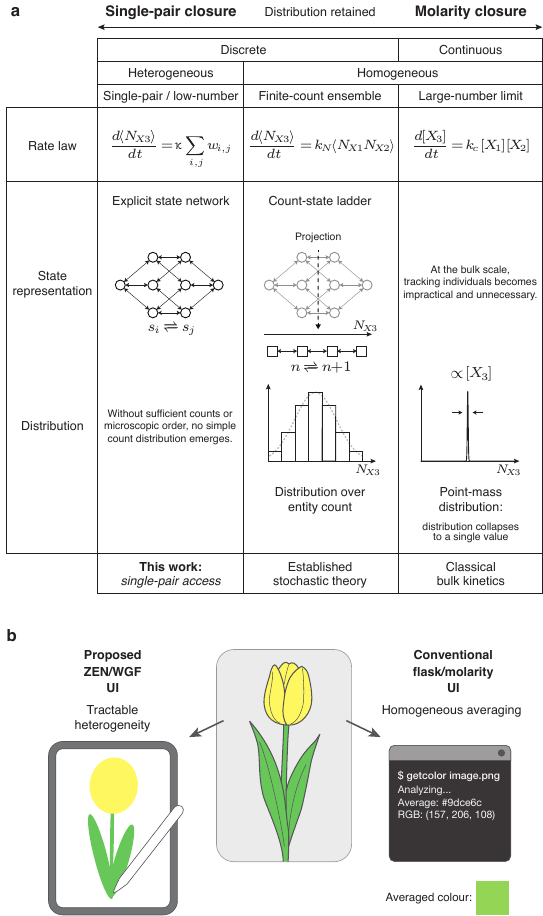}
  \caption{\textbf{Scalar closures of distributional kinetics at single-pair and bulk limits.}
\textbf{a}, Single-pair access provides a complementary scalar closure for distributional kinetics.
Concentration is the familiar bulk closure: in homogeneous large-number systems, individual entities need not be tracked, the distribution collapses to a point-mass concentration, and classical mass-action kinetics is recovered.
At the opposite single-pair limit, local heterogeneity remains explicit and no simple count distribution emerges, yet the edge-weight framework gives a tractable scalar variable through \(C_{\mathrm{eff}}\).
Between these limits, established stochastic theory retains a distribution over entity count.
The single-pair closure makes distribution-dependent kinetic effects experimentally tractable by preserving information that bulk molarity would average away.
\textbf{b}, A scientific-UI analogy contrasts tractable heterogeneity with homogeneous averaging.
The proposed ZEN/WGF interface makes heterogeneous nanoscale arrangements experimentally tractable, with the capacity to encode the positions and spreads of interacting distributions, as suggested by the structured tulip image.
The conventional flask/molarity interface returns a homogeneous scalar readout, analogous to averaging the same image to one colour.
Molarity remains powerful when averaging is appropriate, but it becomes a bottleneck when spatial heterogeneity controls nanoscale molecular interactions.
By giving direct experimental access to that heterogeneity, the ZEN/WGF interface provides a new scientific UI for structured-space molecular interactions.}
  \label{fig6}
  \end{center}
\end{figure}

Beyond defining scalar limits, the interface makes structured-space kinetics experimentally navigable.
By separating local molecular proximity from bulk abundance, it turns nanoscale heterogeneity from a post hoc explanation for assay mismatch into a variable that can be designed and tested.
Structured space matters across molecular sciences, and is especially central in biology, where molecules operate in membranes, pores, condensates, scaffolds and confined volumes.
Dynamic remodelling of organelles and condensates may therefore tune molecular encounters, linking cellular form to biochemical function.
Spatial distributions that share the same $w$ should yield equivalent interaction outcomes for a given molecular pair, allowing simplified \textit{in vitro} geometries to probe architecture-dependent behaviours in complex cellular environments.
Because $w$ does not require an ensemble-averaged concentration as its input, the same UI can bring modern kinetic theory, molecular-resolution simulations and experiments to bear on the \textit{in vitro}--\textit{in vivo} gap.

A companion study already demonstrates the practical value of this control by quantifying weak interactions beyond the limits of conventional assays and extending the same measurement principle to drug-screening applications\cite{CompanionPaper}.
The same distributional logic should extend to protein folding\cite{Onuchic1997}, condensate dynamics\cite{Banani2017}, signalling assemblies\cite{Good2011}, and proximity-driven processes from multivalent binding to transport in confined pores\cite{Gounder2013}.
These are precisely the settings in which bulk measurements may have systematically misread, or entirely missed, the geometry-dependent interactions that matter most\cite{RivasMinton2016}.
The flask-and-molarity UI is like mixing paints: it gives one uniform colour, but erases the brushstrokes that made the image (Fig.~6c).
The ZEN/WGF interface keeps those strokes visible by encoding the $w$-defining placement and spread of molecular probability densities, so $w$-matched simplified geometries can represent more complex spatial arrangements (Fig.~6c).
The resulting spatial representation becomes a shared object for theory and experiment, shifting the readout from a final mean to the arrangement that generated it.
Molecular science now has reason to look past the average.

\bibliography{ref}

\backmatter

\begin{itemize}
 \item[{\small Acknowledgements}] This work was supported by JSPS KAKENHI (grant No. JP23K26777 to D. F.), AMED (grant No. 22am0401020h0004 to D. F.), JST ACT-X (grant No. JPMJAX20BK to M. Y.) and JST FOREST (grant Nos. JPMJFR203R to D. F. and JPMJFR220E to M. Y.). D. F. was additionally supported by JST-Mirai (grant No. JPMJMI22H5), JKA (KEIRIN RACE promotion funds), the Inamori Foundation (InaRIS Fellowship) and the Asian Young Scientist Fellowship.

 \item[{\small Author contributions}] F. M. and D. F. developed the theoretical framework. M. Y. performed the experimental validation. S. I. contributed to discussion and reviewed the theoretical analysis. D. F. conceived, supervised and directed the project. All authors critically read and approved the manuscript.

 \item[{\small Competing interests}] The authors declare no competing interests arising from the theoretical framework presented in this work. The framework was validated using the ZEN experimental platform described in the companion experimental paper\cite{CompanionPaper}; Kyoto University is the applicant for a pending international patent application covering that platform (WO 2025/079662 A1), on which D. F., M. Y. and F. M. are named inventors. D. F. is a co-founder of SeedFairing Institute, a Japanese non-profit general incorporated association established to facilitate dissemination of that technology. S. I. declares no competing interests.

 \item[{\small Data and code availability}] Code for the Bayesian $K_{\mathrm{d}}$ MCMC response-model fitting used in this work and its Supplementary Information is available at \url{https://github.com/FujitaG/zen-response-model}, and code for the tethered-sphere Monte Carlo overlap calculator at \url{https://github.com/FujitaG/zen-spatial-overlap}. Both are archived at Zenodo (\url{https://doi.org/10.5281/zenodo.21777959} and \url{https://doi.org/10.5281/zenodo.21777964}, respectively). Experimental data referenced for validation are available in the companion experimental paper\cite{CompanionPaper} and its Supplementary Information, or from the corresponding author upon reasonable request.

 \item[{\small Use of large language models}] Large language models (Claude, Anthropic; GPT, OpenAI) were used to improve the readability and language of the manuscript, and to assist in writing the analysis code listed above.

 \item[{\small Correspondence}] Correspondence and requests for materials should be addressed to Daishi Fujita~(dfujita@icems.kyoto-u.ac.jp).
\end{itemize}

\bmhead{Supplementary information}

\begin{appendices}
\begin{bibunit}
\section{Introduction of edge weights and reaction rates}
\label{SI_wij}

This section details the implementation of edge weights and reaction rates
using the probabilistic model presented in Eq.~\eqref{eq_weight} as our primary example.
We chose this model for its intuitive form and its compatibility with physical control in the ZEN system.
However, as emphasized in the main text, the UI framework is not tied to
any single model---alternative formulations based on temporal evolution,
energy landscapes, or empirical correlations can equally serve as weights,
provided they satisfy additivity.

We begin with fundamental weight concepts,
proceed through the specific implementation in Eq.~\eqref{eq_weight},
and extend to many-entity and general reaction systems.
We conclude by establishing correspondence with classical theory
in the homogeneous limit and verifying compatibility with experimental parameters.
We show that the framework encompasses established theory and extends
naturally to heterogeneous systems.

\subsection{Two-entity system: Definition of $w_{i,j}$}

The weighted graph framework introduced in the main text expresses reaction rates as a sum of edge weights. In this representation, each node corresponds to an entity and each edge to an entity pair. For their sum to represent the reaction rate, each weight must quantify the reaction propensity, specifically the encounter propensity, of the corresponding entity pair. The development below applies to reactions, binding, and molecular interactions alike; we keep the term reaction rate for $r$, following the main text.

Let $r_{i,j}$ denote the reaction rate contribution from entity pair $(i,j)$. The total reaction rate is then expressed as $r=\sum_{i,j} r_{i,j}$, where the sum runs over reactive pairs. While $r_{i,j}$ itself could serve as the weight, reaction rates depend on multiple parameters including chemical identity and temperature. In contrast, concentration in classical theory is determined solely by volume and entity count, independent of identity or temperature. Following this analogy, the edge weights are defined to depend only on the spatial distribution geometry, thereby providing identity-agnostic parameters of practical utility.

For entities of the same class under uniform global parameters, common factors can be collected into $\kappa$, yielding $r = \kappa \sum_{i,j} w_{i,j}$. Here, entities with different internal states are treated as distinct classes. Consequently, $w_{i,j}$ is determined solely by geometric properties, specifically the spatial distributions of the entities. To provide an explicit definition of $w_{i,j}$, we introduce the following probabilistic model.
The reaction process decomposes into two sequential stages:
\begin{enumerate}
  \item Entities reach a spatial configuration conducive to reaction
  \item The actual reaction event occurs
\end{enumerate}
We denote the first stage as an \textbf{encounter}.

\begin{screen}
\begin{define}
Consider a probabilistic model with repeated independent trials at frequency $k_{\mathrm{enc}}$ (units: $\mathrm{s}^{-1}$).
Let $w_{i,j}$ denote the probability that entities $i$ and $j$ encounter in each trial.
The total encounter rate is then
\begin{equation}
  \nu_{\mathrm{enc}}^{\mathrm{total}} = k_{\mathrm{enc}} \sum_{i,j} w_{i,j}
\end{equation}
Given the conditional probability $p_{\mathrm{evt|enc}}$ that an encounter results in reaction,
the total reaction rate becomes
\begin{equation}
  \begin{aligned}
r &= p_{\mathrm{evt|enc}} \; \nu_{\mathrm{enc}}^{\mathrm{total}} \\
&= p_{\mathrm{evt|enc}} k_{\mathrm{enc}} \sum_{i,j} w_{i,j}
  \end{aligned}
\end{equation}
\end{define}
\end{screen}

Setting $k_{\mathrm{evt}} := p_{\mathrm{evt|enc}} k_{\mathrm{enc}}$ yields $r = k_{\mathrm{evt}} \sum_{i,j} w_{i,j}$, confirming that this definition of $w_{i,j}$ satisfies the additivity condition. Since the weights need only satisfy the additivity condition $r = \kappa \sum_{i,j} w_{i,j}$ stated in the main text, the probabilistic interpretation presented here is not mandatory but serves as one illustrative implementation.

Using this probabilistic model requires attention to its intrinsic timescales. The model involves three characteristic times: the encounter trial time $\tau_{\mathrm{enc}} = k_{\mathrm{enc}}^{-1}$, the characteristic time for distribution evolution $\tau_{f}$, and the observational time window $t_{\mathrm{window}}$. The validity of the framework requires the ordering $\tau_{\mathrm{enc}} \ll t_{\mathrm{window}} \ll \tau_{f}$: the window must span many encounter trials (\(\tau_{\mathrm{enc}} \ll t_{\mathrm{window}}\)) for the encounter probability to be well sampled, yet stay short enough that the distribution is quasi-static (\(t_{\mathrm{window}} \ll \tau_{f}\)).

\subsection{Encounter probability in two-entity systems}
\label{2particles}

Consider two entities following a joint probability distribution $f(\bm{x}_1,\bm{x}_2;t)$ in the probabilistic model of the previous section. Computing the probability that they reach a configuration conducive to reaction first requires a mathematical description of such configurations.

Let $\mathrm{Enc}(\mathrm{encounter};\bm{x}_1,\bm{x}_2)$ denote an encounter criterion function that assigns a value in $[0,1]$ to each configuration, representing how conducive $(\bm{x}_1,\bm{x}_2)$ is to reaction.\footnote{The simplest choice is binary, returning 1 for reactive configurations and 0 otherwise. More general graded forms and idealized distributional limits (e.g., delta-function representations) will be introduced later; see also Appendix \ref{ap_Enc_extention}.} The encounter probability, the probability that two entities meet, is then given by:
\begin{equation}
p_{\mathrm{enc}}(\mathrm{encounter};t) = \int d^3 \bm{x}_1 d^3 \bm{x}_2 f(\bm{x}_1,\bm{x}_2;t) \mathrm{Enc}(\mathrm{encounter};\bm{x}_1, \bm{x}_2).
\label{pEnc_en}
\end{equation}
Here, $p$ necessarily satisfies $0 \leq p \leq 1$:
\begin{equation}
\begin{aligned}
p_{\mathrm{enc}}(\mathrm{encounter};t) &= \int d^3 \bm{x}_1 d^3 \bm{x}_2 f(\bm{x}_1,\bm{x}_2;t) \mathrm{Enc}(\mathrm{encounter};\bm{x}_1, \bm{x}_2) \\
&\le \int d^3 \bm{x}_1 d^3 \bm{x}_2 f(\bm{x}_1,\bm{x}_2;t) = 1.
\end{aligned}
\end{equation}

As shown below, this $p_{\mathrm{enc}}(\mathrm{encounter};t)$ sets the reaction probability of a given entity pair, and therefore serves as the weight in the weighted graph of the previous section. For a two-entity system a graph is unnecessary, so we keep the $p_{\mathrm{enc}}(\mathrm{encounter};t)$ notation directly.

We assume the encounter probability is given by Eq.~\eqref{pEnc_en}.
\begin{screen}
  \begin{define}
    For two entities following a probability distribution $f(\bm{x}_1, \bm{x}_2; t)$, the encounter probability is expressed using an \textbf{encounter criterion function} $\mathrm{Enc}(\mathrm{encounter};\bm{x}_1, \bm{x}_2)$ as:
\begin{equation}
p_{\mathrm{enc}}(\mathrm{encounter};t) = \int d^3 \bm{x}_1 d^3 \bm{x}_2 f(\bm{x}_1,\bm{x}_2;t) \mathrm{Enc}(\mathrm{encounter};\bm{x}_1, \bm{x}_2).
\end{equation}

Furthermore, the encounter frequency $\nu_{\mathrm{enc}}$ per unit time is assumed to be proportional to the encounter probability $p_{\mathrm{enc}}(\mathrm{encounter};t)$:
\begin{equation}
\begin{aligned}
\nu_{\mathrm{enc}} &= k_{\mathrm{enc}}p_{\mathrm{enc}}(\mathrm{encounter};t) \\
&= k_{\mathrm{enc}} \int d^3 \bm{x}_1 d^3 \bm{x}_2 f(\bm{x}_1,\bm{x}_2;t) \mathrm{Enc}(\mathrm{encounter};\bm{x}_1, \bm{x}_2),
\end{aligned}
\end{equation}
where $k_{\mathrm{enc}}$ denotes the proportionality constant, distinguished from the reaction rate constant $k$.
    \label{hypo1}
  \end{define}
\end{screen}
The proportionality constant $k_{\mathrm{enc}}$ also depends on the mean entity velocity.

\begin{ex}
Consider a one-dimensional system where the encounter criterion function is given by a rectangular function:
\begin{equation}
\mathrm{Enc}(\mathrm{encounter};x_1, x_2) = \mathrm{rect}((x_1-x_2)/a_{\mathrm{enc}}),
\end{equation}
where
\begin{equation}
\mathrm{rect}(x) = \left\{
  \begin{array}{ll}
  0 & (|x|>1/2) \\
  1/2 & (|x|=1/2) \\
  1 & (|x|<1/2)
  \end{array}
\right. .
\end{equation}
For clarity in later discussions, assume $x_1$ and $x_2$ are independent, yielding $f(x_1,x_2;t) = f_1(x_1;t)f_2(x_2;t)$.
When the spatial variation of $f_1(x_1;t)$ and $f_2(x_2;t)$ is small compared to $a_{\mathrm{enc}}$, the rectangular function can be approximated by a delta function:
\begin{equation}
\mathrm{Enc}(\mathrm{encounter};x_1, x_2)\simeq a_{\mathrm{enc}}\; \delta(x_1-x_2),
\label{def_enc_delta}
\end{equation}
giving:
\begin{eqnarray}
p_{\mathrm{enc}}(\mathrm{encounter};t) &\simeq& a_{\mathrm{enc}} \int dx_1 dx_2 \; f_1(x_1;t) f_2(x_2;t) \delta(x_1-x_2).
\end{eqnarray}
Alternatively, for later generalization:
\begin{equation}
p \simeq a_{\mathrm{enc}} \int dx dx_1 dx_2 \; f_1(x_1;t) f_2(x_2;t) \delta (x-x_1) \delta (x-x_2).
\end{equation}

Dimensional analysis of $p_{\mathrm{enc}}(\mathrm{encounter};x_1,x_2)$ yields:
\begin{equation}
  \begin{aligned}
p_{\mathrm{enc}}&(\mathrm{encounter};x_1,x_2,t) \\
&= a_{\mathrm{enc}}[L] \int dx dx_1 dx_2 \; f_1(x_1;t)[L^{-1}] f_2(x_2;t)[L^{-1}] \delta (x-x_1) \delta (x-x_2)[L^{-1}].
  \end{aligned}
\end{equation}
Therefore:
\begin{equation}
[p_{\mathrm{enc}}(\mathrm{encounter};x_1,x_2)] = [L^3 \cdot L^{-1} \cdot L^{-1} \cdot L^{-1}] = [1].
\end{equation}
This dimensionless result is consistent with our definition of $p_{\mathrm{enc}}(\mathrm{encounter};x_1,x_2)$ as a probability in Definition~\ref{hypo1}.
\end{ex}

\subsection{Two-component many-entity systems}

We now extend the two-entity treatment of Section~\ref{2particles} to many entities.

Let the system contain $N_1$ entities of class $X_1$ and $N_2$ entities of class $X_2$. Each entity is labelled as $X_{1,i} \; (i=1,\dots ,N_1)$ and $X_{2,j} \; (j=1,\dots ,N_2)$, respectively.

Each entity has its own random variable, and together they follow the joint probability density $f(x_{1,1},x_{1,2},\dots ,x_{1,N_1},x_{2,1},x_{2,2},\dots ,x_{2,N_2};t)$.
We approximate the total encounter frequency as the sum of pairwise encounter frequencies:
\begin{screen}
  \begin{define}
    The total encounter frequency between entities of class 1 and class 2 in the system equals the sum of two-entity encounter frequencies
\begin{equation}
\begin{aligned}
\nu_{\mathrm{enc}}(X_1,X_2;t) &= k_{\mathrm{enc}}p_{\mathrm{enc}}(\mathrm{encounter};t) \\
&= k_{\mathrm{enc}}\int d^3 \bm{x}_1 d^3 \bm{x}_2 f(\bm{x}_1, \bm{x}_2;t) \mathrm{Enc}(\mathrm{encounter};\bm{x}_1, \bm{x}_2 ,t),
\end{aligned}
\label{pair}
\end{equation}
taken over all entity pairs:
\begin{equation}
\begin{aligned}
\nu_{\mathrm{enc}}^{\mathrm{total}} &= \sum_{i=1}^{N_1} \sum_{j=1}^{N_2} \nu_{\mathrm{enc}}(X_{1,i},X_{2,j};t) \\
&= k_{\mathrm{enc}} \sum_{i=1}^{N_1} \sum_{j=1}^{N_2} p_{\mathrm{enc}}^{i,j}(\mathrm{encounter}; t) \\
&= k_{\mathrm{enc}} \sum_{i=1}^{N_1} \sum_{j=1}^{N_2} \int d^3 \bm{x}_{1,i} d^3 \bm{x}_{2,j} \; f_{i,j}(\bm{x}_{1,i},\bm{x}_{2,j};t) \mathrm{Enc}(\mathrm{encounter};\bm{x}_{1,i}, \bm{x}_{2,j},t).
\end{aligned}
\label{nu}
\end{equation}
Here, $f_{i,j}(\bm{x}_{1,i},\bm{x}_{2,j};t)$ denotes the marginalization of the full joint probability distribution $f(x_{1,1},x_{1,2},\dots ,x_{1,N_1},x_{2,1},x_{2,2},\dots ,x_{2,N_2};t)$ over all variables except $\bm{x}_{1,i}$ and $\bm{x}_{2,j}$.
    \label{hypo2}
  \end{define}
\end{screen}

As noted after Definition~\ref{hypo1}, the proportionality constant $k_{\mathrm{enc}}$ is taken to be common across all entity pairs. Although $k_{\mathrm{enc}}$ depends on the mean entity velocities, this assumption is reasonable in solution or crowded biological environments, where local isothermal conditions make it independent of the particular pair. When this does not hold, $k_{\mathrm{enc}}$ must be treated separately for each pair.

Removing the common factor $k_{\mathrm{enc}}$ from the pairwise encounter frequency $\nu_{\mathrm{enc}}(X_{1,i},X_{2,j};t)$, we define the remaining probability $p_{\mathrm{enc}}^{i,j}(\mathrm{encounter};t)$ as the edge weight in the weighted graph for many-entity systems:

\begin{screen}
  \begin{define}[Definition of weight]
\begin{equation}
\begin{aligned}
w(X_{1,i}, X_{2,j}) &:= p_{\mathrm{enc}}^{i,j}(\mathrm{encounter}; t) \\
&=\int d^3 \bm{x}_{1,i} d^3 \bm{x}_{2,j} \; f_{i,j}(\bm{x}_{1,i}, \bm{x}_{2,j};t) \mathrm{Enc}(\mathrm{encounter};\bm{x}_{1,i}, \bm{x}_{2,j},t)
\end{aligned}
\label{def_weight}
\end{equation}
\end{define}
\end{screen}

The joint distribution is more general and, in principle, the more rigorous choice. For the user-centred interface emphasized in the main text, however, the independent treatment carries a decisive practical advantage: it keeps the weight tractable and turns $w$ into a control parameter the experimenter can set and read directly. We therefore adopt the independent distribution as our working choice wherever it holds as a sufficient approximation, and reserve the joint distribution for cases where correlations cannot be neglected. The weight then becomes:

\begin{screen}
\begin{define}[Definition of weight (independent case)]
\begin{equation}
w(X_{1,i}, X_{2,j}) =\int d^3 \bm{x}_1 d^3 \bm{x}_2 \; f_{1,i}(\bm{x}_1;t) f_{2,j}(\bm{x}_2;t) \mathrm{Enc}(\mathrm{encounter};\bm{x}_1, \bm{x}_2,t)
\label{def_weight_indep}
\end{equation}
\end{define}
\end{screen}

We keep this independence assumption through the discussion and generalizations that follow, returning to the general distribution where it is needed.

Having expressed the encounter frequency, we turn to the reaction frequency. When entities of class 1 and class 2 react, the reaction rate is proportional to Eq.~\eqref{nu}.

Let $p_{\mathrm{evt|enc}}$ denote the reaction probability when entities of class 1 and class 2 encounter in the system, and define $k_{\mathrm{evt}}:=p_{\mathrm{evt|enc}} \cdot k_{\mathrm{enc}}$. The reaction rate $r$ is then given by:
\begin{equation}
\begin{aligned}
r &= p_{\mathrm{evt|enc}} \cdot \nu_{\mathrm{enc}}^{\mathrm{total}} \\
&= k_{\mathrm{evt}} \sum_{i=1}^{N_1} \sum_{j=1}^{N_2} p_{\mathrm{enc}}^{i,j}(\mathrm{encounter};t) \nonumber \\
&= k_{\mathrm{evt}} \sum_{i=1}^{N_1} \sum_{j=1}^{N_2} \int d^3 \bm{x}_{1,i} d^3 \bm{x}_{2,j} f_{i,j}(\bm{x}_{1,i}, \bm{x}_{2,j};t) \mathrm{Enc}(\mathrm{encounter};\bm{x}_{1,i}, \bm{x}_{2,j},t) \\
&= k_{\mathrm{evt}} \sum_{i=1}^{N_1} \sum_{j=1}^{N_2} w(X_{1,i}, X_{2,j}).
\end{aligned}
\label{psum}
\end{equation}

Therefore, the system's overall reaction rate is obtained by multiplying the sum of weights by $k_{\mathrm{evt}}$.

The same result follows from summing the pairwise reaction-rate contributions over all entity pairs. Defining the pairwise reaction rate for entities $X_{1,i}$ and $X_{2,j}$ as $r(X_{1,i}, X_{2,j}) := k_{\mathrm{evt}} w(X_{1,i}, X_{2,j})$, the total reaction rate becomes:
\begin{equation}
\begin{aligned}
r &= \sum_{i=1}^{N_1} \sum_{j=1}^{N_2} r(X_{1,i}, X_{2,j}) \\
&= k_{\mathrm{evt}} \sum_{i=1}^{N_1} \sum_{j=1}^{N_2} w(X_{1,i}, X_{2,j}).
\end{aligned}
\end{equation}

\subsection{Generalized Elementary Reaction Law of Mass Action}

We now generalize the two-component result to a general reaction scheme:
    \begin{equation}
    \sum_{{\ell}=1}^{m} s_{\ell} X_{\ell} \ce{->} \sum_{{\ell}=1}^{m} r_{\ell} X_{\ell}.
    \end{equation}
Here, $m$ is the total number of entity classes and $\ell$ labels the class.
Each entity class $X_{\ell}$ has $N_{\ell}$ copies in the system.

\begin{screen}
  \begin{define}[Encounter probability for generalized coordinates]
Let the position of an entity of class $X_{\ell}$ be given by generalized coordinates $\bm{q}_{\ell}$ of dimension $D_{\ell}$. Let $\bm{q}_{{\ell},k}$ denote the coordinates of the $k$-th entity $X_{\ell,k}$, with each entity following the probability density $f_{{\ell},k}(\bm{q}_{{\ell},k})$. The encounter probability $p_{\mathrm{enc}}(\mathrm{encounter};t)$ for the entities involved in the reaction is a functional of the distributions $\{f_{{\ell},k}\}$, defined through the \textbf{encounter criterion function} $\mathrm{Enc}(\mathrm{encounter};\{\bm{q}_{{\ell},k}\}):=\mathrm{Enc}(\mathrm{encounter};\bm{q}_{1,1},\dots ,\bm{q}_{1,s_1},\bm{q}_{2,1},\dots ,\bm{q}_{2,s_2},\dots \dots ,\bm{q}_{m,1},\dots ,\bm{q}_{m,s_{m}})$:
\begin{equation}
p_{\mathrm{enc}} (\mathrm{encounter};\{f_{{\ell},k}\},t) = \int \prod_{{\ell}=1}^{m} \prod_{k=1}^{s_{\ell}} d^{D_{\ell}}\bm{q}_{{\ell},k} \, \mathrm{Enc}(\{\bm{q}_{{\ell},k}\}) \prod_{{\ell}=1}^{m} \prod_{k=1}^{s_{\ell}} f_{{\ell},k} (\bm{q}_{{\ell},k}).
\end{equation}
The encounter criterion function is assumed nonnegative and to yield a well-defined probability through Eq.~\eqref{nu_general}. In regularized forms it takes values in $[0,1]$, and distributional limits are also allowed.

The encounter frequency $\nu_{\mathrm{enc}}(\{X_{{\ell},k}\};t)$ is assumed proportional to the encounter probability $p_{\mathrm{enc}}(\mathrm{encounter};t)$:
\begin{equation}
\begin{aligned}
\nu_{\mathrm{enc}} (\{X_{{\ell},k}\};t) &:= k_{\mathrm{enc}}p_{\mathrm{enc}}(\mathrm{encounter};\{f_{{\ell},k}\},t) \\
&= k_{\mathrm{enc}} \int \prod_{{\ell}=1}^{m} \prod_{k=1}^{s_{\ell}} d^{D_{\ell}}\bm{q}_{{\ell},k} \, \mathrm{Enc}(\{\bm{q}_{{\ell},k}\}) \prod_{{\ell}=1}^{m} \prod_{k=1}^{s_{\ell}} f_{{\ell},k} (\bm{q}_{{\ell},k}).
\end{aligned}
\label{nu_general}
\end{equation}
Here, $k_{\mathrm{enc}}$ denotes the proportionality constant, distinct from the rate constant $k$ in reaction kinetics.
  \end{define}
\end{screen}

The total encounter frequency in the system is the sum of $\nu_{\mathrm{enc}}(\{X_{{\ell},k}\};t)$ from Eq.~\eqref{nu_general} over all entity selections:
\begin{equation}
\begin{aligned}
\nu_{\mathrm{enc}}^{\mathrm{total}} &:= \sum_{\{i_{{\ell},k}\}} \nu_{\mathrm{enc}}(\{X_{{\ell},i_{{\ell},k}}\})(t) \\
&= k_{\mathrm{enc}} \sum_{\{i_{{\ell},k}\}} p_{\mathrm{enc}}(\mathrm{encounter};\{f_{{\ell},i_{{\ell},k}}\},t) \\
&= k_{\mathrm{enc}} \sum_{\{i_{{\ell},k}\}} \int \prod_{{\ell}=1}^{m} \prod_{k=1}^{s_{\ell}} d^{D_{\ell}}\bm{q}_{{\ell},i_{{\ell},k}} \, \mathrm{Enc}(\{\bm{q}_{{\ell},i_{{\ell},k}}\}) \prod_{{\ell}=1}^{m} \prod_{k=1}^{s_{\ell}} f_{{\ell},i_{{\ell},k}} (\bm{q}_{{\ell},i_{{\ell},k}}).
\end{aligned}
\label{nu_general_sum}
\end{equation}
  Here, the summation is taken over all indices satisfying $1 \le i_{1,1} < i_{1,2}<...<i_{1,s_1} \le N_1, 1 \le i_{2,1} < i_{2,2}<...<i_{2,s_2} \le N_2, \dots, 1 \le i_{m,1} < i_{m,2}<...<i_{m,s_m} \le N_{m}$.

Let $p_{\mathrm{evt|enc}}$ denote the reaction probability once the required entities encounter, and define $k_{\mathrm{evt}}:= p_{\mathrm{evt|enc}} \cdot k_{\mathrm{enc}}$. The reaction rate $r$ is then:
  \begin{equation}
  \begin{aligned}
  r &= p_{\mathrm{evt|enc}} \cdot \nu_{\mathrm{enc}}^{\mathrm{total}} \\
  &= k_{\mathrm{evt}} \sum_{\{i_{{\ell},k}\}} p_{\mathrm{enc}}(\mathrm{encounter};\{f_{{\ell},i_{{\ell},k}}\},t) \\
&= k_{\mathrm{evt}} \sum_{\{i_{{\ell},k}\}} \int \prod_{{\ell}=1}^{m} \prod_{k=1}^{s_{\ell}} d^{D_{\ell}}\bm{q}_{{\ell},i_{{\ell},k}} \, \mathrm{Enc}(\{\bm{q}_{{\ell},i_{{\ell},k}}\}) \prod_{{\ell}=1}^{m} \prod_{k=1}^{s_{\ell}} f_{{\ell},i_{{\ell},k}} (\bm{q}_{{\ell},i_{{\ell},k}}).
\end{aligned}
  \label{newLMA_general}
\end{equation}

The quantity $p_{\mathrm{enc}}(\mathrm{encounter};\{f_{{\ell},{k}}\},t)$ here is the generalized weight, denoted $w(\{X_{{\ell},k}\})$.

\begin{screen}
\begin{define}[Generalized weight]
  \begin{equation}
  \begin{aligned}
  w(\{X_{{\ell},k}\}) &:= p_{\mathrm{enc}}(\mathrm{encounter};\{f_{{\ell},{k}}\},t) \\
  &= \int \prod_{{\ell}=1}^{m} \prod_{k=1}^{s_{\ell}} d^{D_{\ell}}\bm{q}_{{\ell},{k}} \, \mathrm{Enc}(\{\bm{q}_{{\ell},{k}}\}) \prod_{{\ell}=1}^{m} \prod_{k=1}^{s_{\ell}} f_{{\ell},{k}} (\bm{q}_{{\ell},{k}})
  \end{aligned}
  \label{gen_weight}
  \end{equation}
\end{define}
\end{screen}

Using this generalized weight, the generalized elementary reaction Law of Mass Action becomes:

\begin{screen}
\begin{define}[Generalized elementary reaction Law of Mass Action (LMA)]
\begin{equation}
r = k_{\mathrm{evt}} \sum_{\{i_{{\ell},k}\}} w(\{X_{{\ell},k}\})
\label{gen_LMA}
\end{equation}
\end{define}
\end{screen}

\begin{ex}[Completely homogeneous probability distribution]
\label{homogeneous}
Consider ordinary Euclidean coordinates $\bm{x}_{{\ell},k}({\ell} = 1, ..., m, k = 1, ..., s_{\ell})$. All entity distributions are uniform in a volume $\Omega$. The encounter criterion function $\mathrm{Enc}$ is defined with a constant $\xi_{\mathrm{enc}}$ (dimensioned so that $\mathrm{Enc}$ is dimensionless) as:
\begin{equation}
\mathrm{Enc}(\{\bm{x}_{{\ell},k}\}) =\xi_{\mathrm{enc}} \int d^3 \bm{x} \prod_{{\ell}=1}^{m} \prod_{k=1}^{s_{\ell}} \delta (\bm{x}_{{\ell},k}-\bm{x}).
\end{equation}
The reaction rate $r$ is then given by:
\begin{eqnarray}
r &=& \Omega k_{\mathrm{evt}} \xi_{\mathrm{enc}} \prod_{{\ell}=1}^{m} \binom{N_{\ell}}{s_{\ell}} \left( \frac{1}{\Omega} \right)^{s_{\ell}} \nonumber \\
&=& \Omega k_{\mathrm{evt}} \xi_{\mathrm{enc}} \prod_{{\ell}=1}^{m} \frac{N_{\ell}!}{(N_{\ell}-s_{\ell})! s_{\ell}! \Omega^{s_{\ell}}}.
\label{LMA_homogeneous}
\end{eqnarray}
This matches the stochastic mass-action formula, up to a factor $1/s_{\ell}!$ that can be absorbed into the reaction rate coefficient.
\end{ex}

\subsection{Relationship between concentration in homogeneous systems and weight in heterogeneous systems}

In conventional chemical kinetics, concentration is the variable, and the reaction rate $r_{\mathrm{c}}$ is given by
\begin{equation}
r_{\mathrm{c}} = k_{\mathrm{c}} \prod_{j=1}^{m} [X_j]^{s_j}.
\label{vc}
\end{equation}
Here $k_{\mathrm{c}}$ is the concentration-based rate constant, which differs from the coefficient in Eq.~\eqref{LMA_homogeneous} by a factor of $\Omega$.
\begin{equation}
[X_j]^{s_j} = \left(\frac{N_j}{\Omega}\right)^{s_j} \simeq \frac{N_j!}{(N_j-s_j)! \Omega^{s_j}}
\end{equation}
Because $r_{\mathrm{c}}$ in Eq.~\eqref{vc} and $r$ in Eq.~\eqref{LMA_homogeneous} differ by a factor of $\Omega$, set
\begin{equation}
k_{\mathrm{c}} = k_{\mathrm{evt}} \xi_{\mathrm{enc}} \prod_{j=1}^{m} \frac{1}{s_j!}.
\label{kc}
\end{equation}
Then Eqs.~\eqref{LMA_homogeneous} and \eqref{vc} become identical. Comparing Eq.~\eqref{gen_LMA} with Eq.~\eqref{LMA_homogeneous} gives
\begin{equation}
\Omega k_{\mathrm{evt}} \xi_{\mathrm{enc}} \prod_{j=1}^{m} \frac{N_j!}{(N_j-s_j)! s_j! \Omega^{s_j}} = k_{\mathrm{evt}} \sum_{\{i_{{\ell},k}\}} w(\{X_{{\ell},k}\})
\end{equation}
.

Although the rate constants differ by a factor of $\xi_{\mathrm{enc}}\prod_{j=1}^{m}(1/s_j!)$, the sum of weights
\begin{equation}
\sum_{\{i_{{\ell},k}\}} w(\{X_{\ell,k}\})
\end{equation}
corresponds to the product of concentrations
\begin{equation}
\Omega \prod_{j=1}^{m} [X_j]^{s_j}.
\end{equation}

\begin{ex}[Treatment of unimolecular reactions]
Consider a unimolecular reaction of the form
\begin{equation}
X_1 \ce{->} X_2.
\end{equation}
When equation (\ref{gen_weight}) is applied to this case, the weight is obtained as
\begin{equation}
\begin{aligned}
w(X_1) &= \int d^3 \bm{x_1} f_{1} (\bm{x}_1) \\
&= 1,
\end{aligned}
\end{equation}
and the reaction rate becomes
\begin{equation}
\begin{aligned}
r &= k_{\mathrm{evt}} \int d^3 \bm{x_1} f_{1} (\bm{x}_1) \\
&= k_{\mathrm{evt}}.
\end{aligned}
\end{equation}
Thus the reaction rate is simply $k_{\mathrm{evt}}$. A unimolecular reaction involves no encounter, so encounter probability does not enter. To mark this case, where a single reactant needs no encounter, we may write $\lambda_{\mathrm{evt}}$ instead of $k_{\mathrm{evt}}$. This notation is also used when determining the equilibrium state.

Temporal uniformity is assumed. The parameter then coincides with the $\lambda$ of a Poisson distribution\cite{Ross1983}. Here, $\lambda$ is used to emphasize that the reaction rate is independent of the weight $w$.
\end{ex}

\subsection{Treatment of two-component homogeneous systems and compatibility with $K_d$}

To relate $k_{\mathrm{evt}}$ and $\lambda_{\mathrm{evt}}$ in two-component systems to the standard dissociation constant $K_d$, consider the reaction $X_1+X_2 \ce{->} X_3$.
When the probability distribution varies little over the spatial extent of the $\mathrm{Enc}$ function, $\mathrm{Enc}$ can be approximated by a delta function:
\begin{equation}
\mathrm{Enc}(\bm{x}_1, \bm{x}_2; t) = \xi_{\mathrm{enc}}^{X1,X2} \delta (\bm{x}_1-\bm{x}_2).
\end{equation}
The weight $w(X_1,X_2)$ is thereby given by
\begin{equation}
\begin{aligned}
w(X_1,X_2) &:= \int_{V} d^3 \bm{x}_{1} d^3 \bm{x}_{2} f_{X1}(\bm{x}_1;t) f_{X2}(\bm{x}_2;t) \mathrm{Enc}(\mathrm{encounter};\bm{x}_1, \bm{x}_2,t) \\
&= \xi_{\mathrm{enc}}^{X1,X2} \int_{V} d^3 \bm{x} f_{X1}(\bm{x};t) f_{X2}(\bm{x};t) \\
&= \xi_{\mathrm{enc}}^{X1,X2} \int_{V} d^3 \bm{x} \frac{1}{V^2} \\
&= \frac{\xi_{\mathrm{enc}}^{X1,X2}}{V}.
\end{aligned}
\end{equation}
The reaction rate $r(X_1+X_2 \rightarrow X_3)$ then becomes
\begin{equation}
\begin{aligned}
r(X_1+X_2 \rightarrow X_3) &= N_{X1} N_{X2} k_{\mathrm{evt}}^{X1+X2\rightarrow X3} w(X_1,X_2) \\
&= \frac{N_{X1} N_{X2} k_{\mathrm{evt}}^{X1+X2\rightarrow X3} \xi_{\mathrm{enc}}^{X1,X2}}{V}.
\end{aligned}
\end{equation}
For the reverse reaction $X_3 \ce{->} X_1+X_2$, the reaction rate is written with the rate constant $\lambda_{\mathrm{evt}}^{X3 \rightarrow X1+X2}$ as
\begin{equation}
r(X_3 \rightarrow X_1+X_2) = N_{X3} \lambda_{\mathrm{evt}}^{X3 \rightarrow X1+X2}.
\end{equation}

At equilibrium, these two reaction rates balance, yielding
\begin{equation}
\frac{N_{X1} N_{X2} k_{\mathrm{evt}}^{X1+X2\rightarrow X3} \xi_{\mathrm{enc}}^{X1,X2}}{V} = N_{X3} \lambda_{\mathrm{evt}}^{X3 \rightarrow X1+X2}.
\end{equation}
Rearranging this expression gives
\begin{equation}
\cfrac{\cfrac{N_{X1}}{N_A V} \cdot \cfrac{N_{X2}}{N_A V}}{\cfrac{N_{X3}}{N_A V}} = \cfrac{\lambda_{\mathrm{evt}}^{X3 \rightarrow X1+X2}}{N_A k_{\mathrm{evt}}^{X1+X2\rightarrow X3} \xi_{\mathrm{enc}}^{X1,X2}},
\end{equation}
where $N_A$ denotes Avogadro's constant. Since the left-hand side is the dissociation constant $K_d$,
\begin{equation}
K_d^{X1,X2} = \frac{[X_1][X_2]}{[X_3]},
\end{equation}
where $[X_i] = N_{Xi}/(N_A V)$. Therefore, the relation
\begin{equation}
K_d^{X1,X2} = \frac{\lambda_{\mathrm{evt}}^{X3 \rightarrow X1+X2}}{N_A k_{\mathrm{evt}}^{X1+X2\rightarrow X3} \xi_{\mathrm{enc}}^{X1,X2}}
\end{equation}
is obtained, or equivalently
\begin{equation}
\frac{k_{\mathrm{evt}}^{X1+X2\rightarrow X3}}{\lambda_{\mathrm{evt}}^{X3 \rightarrow X1+X2}} = \frac{1}{N_A \xi_{\mathrm{enc}}^{X1,X2} K_d^{X1,X2}}.
\label{Kd}
\end{equation}

\section{Extensibility of the Enc function}
\label{ap_Enc_extention}

In the examples so far, the $\mathrm{Enc}$ function has taken delta-function forms:
\begin{equation}
\mathrm{Enc}(\mathrm{encounter}; \bm{x}_1, \bm{x}_2) := \xi_{\mathrm{enc}}^{X1,X2} \delta (\bm{x}_1-\bm{x}_2),
\end{equation}
and
\begin{equation}
\mathrm{Enc}(\{\bm{x}_{{\ell},k}\}) =\xi_{\mathrm{enc}} \int d^3 \bm{x} \prod_{{\ell}=1}^{m} \prod_{k=1}^{s_{\ell}} \delta (\bm{x}_{{\ell},k}-\bm{x}).
\end{equation}
Delta functions, however, are not generally required. For example, when the interaction volume $\xi_{\mathrm{enc}}^{X1,X2}$ is not small compared with the total system volume, the delta-function approximation for $\mathrm{Enc}$ breaks down. In such cases, the weight must be computed from its defining equations, Eqs.~\eqref{def_weight} and \eqref{gen_weight}.

Consider whether a relation analogous to Eq.~\eqref{Kd}, connecting $k_{\mathrm{evt}}$, $\lambda_{\mathrm{evt}}$ and $K_d$, can still be established when $\mathrm{Enc}$ cannot be approximated by a delta function. The correspondence with $K_d$ was derived for a homogeneous system, so we take uniform distributions $f_{X1}=f_{X2}=1/V$ and let $\mathrm{Enc}$ be time-independent and translation-symmetric, $\mathrm{Enc}(\bm{x}_1, \bm{x}_2; t) = \mathrm{Enc}(\bm{x}_1-\bm{x}_2)$. Defining
\begin{equation}
\xi_{\mathrm{enc}} := \int d\bm{x}\, \mathrm{Enc}(\bm{x}),
\label{xi}
\end{equation}
which has dimensions of volume, the weight in Eq.~\eqref{def_weight} reduces in the bulk limit to
\begin{equation}
w(X_1,X_2) = \frac{1}{V^2}\int_V d\bm{x}_1\, d\bm{x}_2\, \mathrm{Enc}(\bm{x}_1-\bm{x}_2) = \frac{\xi_{\mathrm{enc}}}{V},
\end{equation}
the same form found for the delta-function case with $\xi_{\mathrm{enc}}^{X1,X2}$. Thus $\xi_{\mathrm{enc}}$ defined by Eq.~\eqref{xi} plays the role of $\xi_{\mathrm{enc}}^{X1,X2}$, and Eq.~\eqref{Kd} carries over unchanged.

Although a binary (indicator) form of $\mathrm{Enc}$ has often been used for simplicity, this restriction is not essential. More generally, $\mathrm{Enc}$ may be taken as a continuous function with values in $[0,1]$, whose magnitude encodes how conducive a configuration is to reaction. Idealized distributional limits, such as delta functions, can serve as sharp approximations where appropriate. This generalization lets a single framework treat diverse interactions, including macromolecular and intramolecular reactions.

\section{Monte Carlo spatial-overlap calculations for Figure 4}
\label{sec:fig4_overlap_simulation}

\subsection{Geometry and orientation sampling}

The spatial distributions and overlap estimates in Fig.~4 were calculated with
a custom Monte Carlo orientation-sampling program. The nanocavity boundary was
represented by 40 circular exclusion sites on a two-dimensional lattice. LgBiT
and SmBiT or DarkBiT were represented as tethered spheres; molecular shape,
linker elasticity, electrostatic interactions and protein--surface interactions
other than hard geometric exclusion were not included. Table~\ref{tab:fig4b_overlap_parameters}
lists the fixed settings for the representative LgBiT--SmBiT geometry shown in
Fig.~4b.

\begin{table}[h!]
\centering
\caption{Parameters used for the representative spatial-density calculation in
Fig.~4b. All lengths are in nanometres.}
\label{tab:fig4b_overlap_parameters}
\begin{tabular}{ll}
\toprule
Parameter & Value \\
\midrule
Nanocavity lattice spacing & \(2.65\) \\
Boundary-site radius & \(1.325\) \\
LgBiT sphere radius & \(3.25\) \\
SmBiT/DarkBiT sphere radius & \(2.25\) \\
LgBiT anchor & \((13.25,14.575)\) \\
SmBiT anchor & \((13.25,1.325)\) \\
Nominal LgBiT and SmBiT linker lengths & \(4.2,\,4.2\) \\
Geometric length offsets & \(1.5,\,1.5\) \\
Simulation bounds & \(29.15\times18.55\times21.40\) \\
Grid size & \(100\times64\times73\) voxels \\
Orientation samples & \(1000\) per \(z\)-plane and component \\
Collision tolerance & \(10^{-6}\) \\
\bottomrule
\end{tabular}
\end{table}

At each reactive-site grid point, the program sampled orientation vectors
uniformly on the surface of the corresponding protein sphere. A configuration
was rejected if the sphere intersected the nanocavity boundary or failed the
linker-reach condition
\begin{equation}
  \ell-R_k+R_k\theta \leq L_k+\Delta L_k,
  \qquad
  \theta=\cos^{-1}\!\left(
    \frac{(\bm o-\bm a_k)\cdot\bm u}{R_k\ell}
  \right),
  \qquad
  \ell=\lVert\bm o-\bm a_k\rVert,
  \label{eq:fig4b_reach_condition}
\end{equation}
where \(\bm a_k\), \(L_k\), \(\Delta L_k\), \(R_k\), \(\bm u\) and
\(\bm o\) denote the anchor position, nominal linker length, length offset,
protein radius, sampled orientation vector and sphere centre, respectively.
Accepted configurations were counted by voxel. The maps in Fig.~4b are
projections of these counts along the \(z\) direction; the overlap calculation
used the full three-dimensional densities.

\subsection{Overlap estimator}

For accepted-configuration count \(n_{k,v}\) of component \(k\) in voxel \(v\),
we defined the discrete voxel probability
\begin{equation}
  p_{k,v}=
  \frac{n_{k,v}}{\sum_{v'}n_{k,v'}}.
  \label{eq:fig4b_voxel_probability}
\end{equation}
Taking one voxel of volume \(\Delta V=\Delta x\,\Delta y\,\Delta z\) as the
discrete encounter cell, the same-voxel encounter weight was
\begin{equation}
  w_{\Delta V}
  =
  \sum_v p_{1,v}p_{2,v}.
  \label{eq:fig4b_discrete_weight}
\end{equation}
The corresponding normalized spatial density was
\(f_{k,v}=p_{k,v}/\Delta V\), and the position-based effective-concentration
estimate was
\begin{equation}
  \begin{aligned}
  c_{\mathrm{eff}}^{\mathrm{MC}}
  &=
  \frac{10^{-3}\times10^{27}}{N_A\Delta V}\,
  w_{\Delta V} \\
  &=
  \frac{10^{-3}\times10^{27}}{N_A}\,
  \Delta V\sum_v f_{1,v}f_{2,v},
  \end{aligned}
  \label{eq:fig4b_overlap_estimator}
\end{equation}
where the prefactor converts an inverse cubic nanometre to molar units. Thus
the same overlap calculation gives both the dimensionless discrete encounter
weight \(w_{\Delta V}\) and its molar representation
\(c_{\mathrm{eff}}^{\mathrm{MC}}\). The implementation records these as
\texttt{discrete\_weight} and \texttt{effective\_concentration\_M}.

\subsection{Binding-fraction comparison in Figure 4c}

For Fig.~4c, the Monte Carlo estimates for the LgBiT--SmBiT and
LgBiT--DarkBiT pairs were converted to binding ratios using
\begin{equation}
  \rho_{\mathrm{Sm}}
  =\frac{a\,c_{\mathrm{eff},\mathrm{Sm}}^{\mathrm{MC}}}
        {K_{\mathrm d,\mathrm{Sm}}},
  \qquad
  \rho_{\mathrm{Dark}}
  =\frac{a\,c_{\mathrm{eff},\mathrm{Dark}}^{\mathrm{MC}}}
        {K_{\mathrm d,\mathrm{Dark}}},
  \label{eq:fig4c_binding_ratios}
\end{equation}
with fixed \(K_{\mathrm d,\mathrm{Sm}}=1.9\times10^{-4}\,\mathrm{M}\) and
\(K_{\mathrm d,\mathrm{Dark}}=1.0\times10^{-5}\,\mathrm{M}\). The theoretical
quantity plotted on the horizontal axis was the SmBiT binding-fraction ratio
with and without DarkBiT,
\begin{equation}
  R_{\mathrm{theory}}
  =\frac{1+\rho_{\mathrm{Sm}}}
         {1+\rho_{\mathrm{Sm}}+\rho_{\mathrm{Dark}}}.
  \label{eq:fig4c_response_ratio}
\end{equation}
The vertical-axis value was the corresponding experimental luminescence ratio,
normalized to the no-DarkBiT control for each construct.

The dimensionless factor \(a\) is a common calibration between the positional
overlap estimate and the reactive effective-concentration scale. It may absorb
orientational, internal-state and other global effects not resolved by the
positional model, and is not identified with \(\xi_{\mathrm{enc}}\) or a unique
microscopic mechanism. A single value was fitted jointly across the two
datasets by least squares against the identity line, giving \(a=0.0504\).
The same value was used for all geometries and both interactions. Across the ten
non-reference conditions, the resulting ratios gave \(r=0.994\) on
natural-log-transformed values and
\(\mathrm{RMSE}_{\ln}=0.087\). Because \(a\) was estimated from the observations
shown in Fig.~4c, this comparison is an internal calibration and consistency
test rather than a fully independent prediction.

\subsection{Numerical variability and scope of uncertainty}

We distinguished stochastic Monte Carlo variability from systematic numerical
and model uncertainty. Because sampled orientations contribute to multiple grid
locations, voxel counts are correlated and were not treated as independent
binomial observations. Numerical repeatability was instead assessed by
independently rerunning the complete simulation.

For the representative geometry in Fig.~4b, three independent runs gave
\[
  c_{\mathrm{eff}}^{\mathrm{MC}}
  = 1.1971 \pm 0.0101\,\mathrm{mM},
\]
where the uncertainty is the sample standard deviation across runs
(\(n=3\)). Thus, the run-to-run coefficient of variation was \(0.85\%\) under
the fixed simulation settings.

The central value is supported by an independent route. Fitting the companion
experimental titration series returns a shared effective concentration of
\(0.978\,\mathrm{mM}\)\cite{CompanionPaper}, which agrees with the present
geometric estimate to within a factor of 1.3, although the two were never
constrained to agree.

This standard deviation quantifies stochastic sampling variability alone and
therefore understates the overall uncertainty in
\(c_{\mathrm{eff}}^{\mathrm{MC}}\): the distributions in Fig.~4b and the
resulting values are conditional on the stated geometry and on the
hard-exclusion model. Finite voxel resolution, the
geometric coarse-graining of the proteins and nanocavity, linker and anchor
parameters, and interactions beyond hard exclusion each offer a route for
refinement; a higher-resolution treatment could represent them explicitly and
propagate their contributions into a combined physical interval.

\subsection{Implementation and reproducibility}

The calculation used a Python interface and a C++ core exposed through pybind11,
with OpenMP parallelization. The representative Fig.~4b input is
\texttt{lgbit\_smbit\_main\_4p2x4p2.txt}; the source files, this fixed input and
sampling settings are distributed at
\url{https://github.com/FujitaG/zen-spatial-overlap}. By default the
random-number generators are initialized from \texttt{std::random\_device}, so
repeated runs produce statistically equivalent rather than bitwise-identical
estimates; the values reported here were obtained with that default.

\section{Weight Normalization and Dimensional Analysis}
\label{sec:normalized_weight}

\subsection{Notational Simplification for Eq. \eqref{eq_rate_particle_counts}}
\label{subsec:rate_equation_normalization}

In the main text, we absorbed the encounter-volume parameter $\xi_{\mathrm{enc}}$ into the rate constant $\kappa$ to improve readability and emphasize the $1/V$ scaling. To keep the weights $w_{i,j}$ dimensionless in $r = \kappa \sum_{i,j} w_{i,j}$ (Eq.~\eqref{eq_additivity_axiom}), however, $\xi_{\mathrm{enc}}$ must be separated from the global scale $\kappa$:
\begin{equation}
\frac{d\langle N_{X3} \rangle}{dt} = \kappa \langle N_{X1} \rangle \langle N_{X2} \rangle \overline{w} = \kappa \langle N_{X1} \rangle \langle N_{X2} \rangle \frac{\xi_{\mathrm{enc}}}{V},
\end{equation}
where $\overline{w} = \xi_{\mathrm{enc}}/V$ is the mean interaction weight over all entity pairs.

We now show how the entity-based form (Eq.~\eqref{eq_rate_particle_counts}) becomes the conventional molarity-based rate equation (Eq.~\eqref{eq_rate_molarity_based}). Converting entity counts to concentrations using $[X_i] = \langle N_{X_i} \rangle/(N_A V)$:
\begin{equation}
\begin{aligned}
\frac{d [X_3]}{dt}[(\mathrm{mol/L}) \cdot \mathrm{s}^{-1}] &= \frac{1}{N_A V} [\mathrm{mol/L}] \frac{d \langle N_{X3} \rangle}{dt} [\mathrm{s}^{-1}] \\
&= \frac{1}{N_A V} \kappa \langle N_{X1} \rangle \langle N_{X2} \rangle \frac{\xi_{\mathrm{enc}}[\mathrm{m}^3]}{V} \\
&= \frac{1}{N_A V} \kappa (N_A V)^2 [X_1][X_2] \frac{\xi_{\mathrm{enc}}[\mathrm{m}^3]}{V} \\
&= (1000 N_A \xi_{\mathrm{enc}}) [(\mathrm{mol/L})^{-1}] \kappa [X_1][X_2] [(\mathrm{mol/L})^2 \cdot \mathrm{s}^{-1}] \\
&= (1000 N_A \xi_{\mathrm{enc}}) \kappa [X_1][X_2] [(\mathrm{mol/L}) \cdot \mathrm{s}^{-1}] .
\end{aligned}
\end{equation}
Setting $k_c = 1000 N_A \xi_{\mathrm{enc}} \kappa$ recovers Eq.~\eqref{eq_rate_molarity_based}.

\section{Closed ZEN System}
\label{closed}
Consider a closed system containing one entity each of $X_1$ and $X_2$. Reactions follow $X_1+X_2 \ce{<=>} X_3$, with probability distributions $f_{X1}(\bm{x})$, $f_{X2}(\bm{x})$, $f_{X3}(\bm{x})$ defined for each entity class. This configuration is termed a closed ZEN system.

The entity-number state is represented by
\begin{equation}
\bm{n} := \begin{pmatrix}
N_{X1} \\ N_{X2} \\ N_{X3}
\end{pmatrix}.
\end{equation}
The closed ZEN system admits only two possible entity-number states:
\begin{equation}
\bm{n}_0 := \begin{pmatrix}
1 \\ 1 \\ 0
\end{pmatrix}, \; \bm{n}_1 := \begin{pmatrix}
0 \\ 0 \\ 1
\end{pmatrix}.
\label{particle_state}
\end{equation}

Here we write the interaction volume as $\xi_{\mathrm{enc}}^{X1,X2}$ to allow for class-dependent interaction scales in later generalizations. Under these conditions, the weights are given by
  \begin{equation}
  \begin{aligned}
w(X_1,X_2) &= \xi_{\mathrm{enc}}^{X1,X2} \int d^3\bm{x} f_{X1}(\bm{x}) f_{X2}(\bm{x}) \\
w(X_3) &= \int d^3\bm{x} f_{X3}(\bm{x}) = 1.
  \end{aligned}
  \end{equation}

Let $r(X_1+X_2\rightarrow X_3)$ and $r(X_3 \rightarrow X_1+X_2)$ denote the forward and reverse reaction rates, respectively:
\begin{equation}
\begin{aligned}
r(X_1+X_2\rightarrow X_3) &= k_{\mathrm{evt}}^{X1+X2\rightarrow X3} w(X_1,X_2) \\
r(X_3 \rightarrow X_1+X_2) &= \lambda_{\mathrm{evt}}^{X3\rightarrow X1+X2}.
\end{aligned}
\end{equation}
Here, the forward event-rate constant is $k_{\mathrm{evt}}^{X1+X2\rightarrow X3}$. Because the reverse reaction $X_3\rightarrow X_1+X_2$ has a single reactant and requires no encounter, its rate constant is written $\lambda_{\mathrm{evt}}^{X3\rightarrow X1+X2}$ rather than $k_{\mathrm{evt}}$.

The master equation therefore becomes
\begin{equation}
\begin{aligned}
\frac{d}{dt} P(\bm{n}_0;t) &= -k_{\mathrm{evt}}^{X1+X2\rightarrow X3} w(X_1,X_2) P(\bm{n}_0;t) + \lambda_{\mathrm{evt}}^{X3\rightarrow X1+X2} P(\bm{n}_1;t) \\
\frac{d}{dt} P(\bm{n}_1;t) &= k_{\mathrm{evt}}^{X1+X2\rightarrow X3} w(X_1,X_2) P(\bm{n}_0;t) - \lambda_{\mathrm{evt}}^{X3\rightarrow X1+X2} P(\bm{n}_1;t). \\
\end{aligned}
\end{equation}

The steady-state solutions, denoted $\pi(\bm{n}_0)$ and $\pi(\bm{n}_1)$, satisfy
\begin{eqnarray}
k_{\mathrm{evt}}^{X1+X2\rightarrow X3} w(X_1,X_2) \pi(\bm{n}_0) - \lambda_{\mathrm{evt}}^{X3\rightarrow X1+X2} \pi(\bm{n}_1) = 0 \\
\iff \frac{\pi(\bm{n}_1)}{\pi(\bm{n}_0)} = \frac{k_{\mathrm{evt}}^{X1+X2\rightarrow X3} w(X_1,X_2)}{\lambda_{\mathrm{evt}}^{X3\rightarrow X1+X2}}.
\end{eqnarray}

Expressing this ratio in terms of $K_d$ using Eq.~\eqref{Kd} yields
\begin{equation}
\frac{\pi(\bm{n}_1)}{\pi(\bm{n}_0)} = \frac{w(X_1,X_2)}{N_A \xi_{\mathrm{enc}}^{X1,X2} K_d^{X1,X2}}.
\end{equation}
The normalization condition $\pi(\bm{n}_1)+\pi(\bm{n}_0)=1$ then gives
\begin{equation}
\pi(\bm{n}_1) = \frac{\frac{w(X_1,X_2)}{N_A \xi_{\mathrm{enc}}^{X1,X2} K_d^{X1,X2}}}{1+\frac{w(X_1,X_2)}{N_A \xi_{\mathrm{enc}}^{X1,X2} K_d^{X1,X2}}}.
\label{p1}
\end{equation}

\subsection{Connection to Michaelis--Menten Kinetics}

Defining an effective concentration as
\begin{equation}
c_{\mathrm{eff}}^{X1,X2} := \frac{w(X_1,X_2)}{N_A \xi_{\mathrm{enc}}^{X1,X2}},
\end{equation}
where $c_{\mathrm{eff}}^{X1,X2}$ has dimensions of molar concentration, Eq.~\eqref{p1} becomes
\begin{equation}
\pi(\bm{n}_1) = \frac{c_{\mathrm{eff}}^{X1,X2}}{K_d^{X1,X2}+c_{\mathrm{eff}}^{X1,X2}}.
\end{equation}
This expression exhibits the same functional form as the Michaelis--Menten equation
\begin{equation}
v = \frac{V_{\mathrm{max}}c}{K_d+c}.
\end{equation}
Here $\pi(\bm{n}_1)$ plays the role of the normalized velocity $v/V_{\mathrm{max}}$, and the effective concentration $c_{\mathrm{eff}}^{X1,X2}$ that of the substrate concentration $c$. The dissociation constant $K_d^{X1,X2}$ plays the same role as the Michaelis constant in fixing the half-saturation point.

The parameter $\xi_{\mathrm{enc}}$ can be interpreted as the coefficient arising when $\mathrm{Enc}$ is approximated by a delta function, or more generally defined via Eq.~\eqref{xi}. When $\mathrm{Enc}$ takes the form $\mathrm{Enc}(\bm{x}_1,\bm{x}_2;t) = \mathrm{Enc}(\bm{x}_1-\bm{x}_2)$, the effective concentration becomes
\begin{equation}
c_{\mathrm{eff}}^{X1,X2} = \cfrac{\displaystyle \int d\bm{x}_1 d\bm{x}_2 \; f_{X1}(\bm{x}_1) f_{X2}(\bm{x}_2) \mathrm{Enc}(\bm{x}_1-\bm{x}_2)}{\displaystyle N_A \int d\bm{x} \; \mathrm{Enc}(\bm{x})}.
\end{equation}

\section{Application to the Open ZEN System}
\label{ap_inhibition_characteristics}

In this section, we apply the weighted graph model introduced in
Section~\ref{SI_wij} to an open ZEN system.
After defining the system, we use a hard threshold to treat each cavity as a
single local environment, which we label a stage.
We assume a separation of timescales, with binding within a stage much faster
than exchange of the non-luminescent competitor between the reservoir and the
cavity. Thus the local competitor count stays fixed within each observation
window while the binding equilibrates. The observed response is then obtained by
averaging the fixed-count equilibrium response over the distribution of local
competitor counts.

\subsection{Open ZEN System}

As introduced in the main text, an entity class can be grouped by chemical
identity, internal state, or spatial localization.
In the open ZEN system the same molecule moves between the cavity and the
reservoir. Labelling by class alone would still distinguish the entities in
principle, but a bare list of class indices makes it hard to see at a glance
whether a change reflects a different molecule or only a different location.
We therefore keep the class label for molecular identity and add a separate
label, the stage, for the reaction venue.
The stage label is redundant in principle; we introduce it only as an aid to
readability, so that molecular identity and reaction venue remain easy to tell
apart.

On this convention, the classes correspond to the split-luciferase components:
\begin{description}
\item[$X_1$:] LgBiT
\item[$X_2$:] SmBiT
\item[$X_3$:] LgBiT--SmBiT
\item[$X_4$:] DarkBiT
\item[$X_5$:] LgBiT--DarkBiT
\end{description}

The two stages are the external reservoir \(S_0\) and the ZEN cavity \(S_1\);
the same class in different stages is distinguished by a stage argument, so
that, for example, \(N_{X4}(S_1)\) counts class \(X_4\) in \(S_1\).

Consider many identical ZEN cavities.
Each cavity (\(S_1\)) contains one \(X_1\) and one \(X_2\), which react as
\(X_1+X_2 \ce{<=>} X_3\).
DarkBiT (\(X_4\)) exchanges between \(S_0\) and \(S_1\) and, while in \(S_1\),
reacts as \(X_1+X_4 \ce{<=>} X_5\); this exchange moves a single class between
stages.
We call this configuration the open ZEN system.

Reactivity follows a hard threshold: two entities react only when they occupy
the same cavity (\(S_1\)) and lie within the interaction radius
\(r_{\mathrm{int}}\).
When the cavity is small enough that every candidate occupies the same
interaction region around \(X_1\), all candidates of one class share a single
geometric encounter weight.
The \(X_4\) moieties in a cavity are then interchangeable competitors for
\(X_1\), each carrying the same weight.
The \(X_2\) and \(X_4\) channels differ only through their weights
\(w((X_1,X_2):S_1)\) and \(w((X_1,X_4):S_1)\), or through the binding and
dissociation rate constants.
A cavity therefore behaves as a single local environment, consistent with
treating it as one stage \(S_1\).

Within \(S_1\), the conserved \(\widehat{X}_4\) moiety count is
\[
N_{\widehat{X4}}(S_1) = N_{X4}(S_1) + N_{X5}(S_1).
\]
Fixing \(N_{\widehat{X4}}(S_1)=n\) then gives \(X_1\) one \(X_2\)-derived candidate
and \(n\) competitors derived from \(\widehat{X}_4\).

In the weighted graph model of Section~\ref{SI_wij}, each entity is a vertex and
each reactive pair an edge, with the edge weight set by the geometric encounter
probability. The reaction-event frequency is the sum over reactive edges. When
several competitors share a local environment, they contribute additively, each
through its own weight.

We take the observation window \(\Delta t_{\mathrm{obs}}\) to be shorter than the
timescale on which \(\widehat{X}_4\) moieties exchange between \(S_0\) and \(S_1\),
but longer than the timescale of binding and dissociation
(\(X_1+X_2 \ce{<=>} X_3\) and \(X_1+X_4 \ce{<=>} X_5\)) within \(S_1\):
\[
\tau_{\mathrm{on/off}} \ll \Delta t_{\mathrm{obs}} \ll \tau_{\mathrm{in/out}}.
\]
Under this separation, \(N_{\widehat{X4}}(S_1)=n\) is fixed within a single
window, while the binding and dissociation within \(S_1\) are fast enough to
reach equilibrium at that fixed \(n\).

The response in each window is then the equilibrium response \(R(n)\) conditional
on \(N_{\widehat{X4}}(S_1)=n\). Across many cavities or observation windows,
different values of \(n\) are sampled, so the observed response is
\[
\bar R = \sum_n \pi(n)\,R(n),
\]
where \(\pi(n)\) is the occupancy distribution, i.e.\ the probability of finding
\(n\) \(\widehat{X}_4\) moieties in \(S_1\), distinct from the state probabilities
\(\pi(\bm{n}_0)\) and \(\pi(\bm{n}_1)\) in Section~\ref{closed}.

In what follows, we fix the notation below and then derive the binding
probability conditional on \(N_{\widehat{X4}}(S_1)=n\).

The following notation is employed:
\begin{description}[labelwidth=4cm]
\item[$N_{\widehat{X4}}(S_1)$] Conserved moiety count of \(\widehat{X}_4\) in stage \(S_1\)
\item[$N_{X_i}(S_1)$] Number of entities \(X_i\) in stage \(S_1\)
\item[$r((X_i+X_j \rightarrow X_k):S_1)$] Reaction rate for \(X_i+X_j \rightarrow X_k\) within \(S_1\)
\item[$k_{\mathrm{evt}}^{(X_i+X_j \rightarrow X_k):S_1}$] Intrinsic event rate constant for pair formation within \(S_1\)
\item[$\lambda_{\mathrm{evt}}^{(X_k \rightarrow X_i+X_j):S_1}$] Intrinsic event rate constant for dissociation within \(S_1\)
\item[$w((X_i,X_j):S_1)$] Geometric encounter weight between \(X_i\) and \(X_j\) within \(S_1\)
\item[$\pi(n)$] Occupancy distribution of \(N_{\widehat{X4}}(S_1)\)
\end{description}

From here on we do not write the spatial distributions explicitly; instead, the
geometric encounter weights of Section~\ref{SI_wij} carry the effect of the
local environment. Specifically, \(w((X_1,X_2):S_1)\) and \(w((X_1,X_4):S_1)\) are
the encounter-probability weights for the \(X_1\)--\(X_2\) and \(X_1\)--\(X_4\)
pairs in \(S_1\), and their definitions already contain the positional
distributions of the entities.

With this notation, we derive the binding probability at fixed occupancy. The
conserved count \(N_{\widehat{X4}}(S_1)=N_{X4}(S_1)+N_{X5}(S_1)\) is unchanged by
\(X_1+X_4 \ce{<=>} X_5\), so it stays at a fixed value \(n\) within an observation
window.

The intra-stage reaction rates in \(S_1\) are expressed as
\begin{align}
r((X_1+X_2 \rightarrow X_3):S_1)
&=
N_{X1}(S_1)N_{X2}(S_1)
k_{\mathrm{evt}}^{(X_1+X_2\rightarrow X_3):S_1}
w((X_1,X_2):S_1), \nonumber \\
r((X_3 \rightarrow X_1+X_2):S_1)
&=
N_{X3}(S_1)
\lambda_{\mathrm{evt}}^{(X_3\rightarrow X_1+X_2):S_1}, \nonumber \\
r((X_1+X_4 \rightarrow X_5):S_1)
&=
N_{X1}(S_1)N_{X4}(S_1)
k_{\mathrm{evt}}^{(X_1+X_4\rightarrow X_5):S_1}
w((X_1,X_4):S_1), \nonumber \\
r((X_5 \rightarrow X_1+X_4):S_1)
&=
N_{X5}(S_1)
\lambda_{\mathrm{evt}}^{(X_5\rightarrow X_1+X_4):S_1}.
\label{eq:intra_stage_rates}
\end{align}

At fixed \(N_{\widehat{X4}}(S_1)=n\), we take the unbound state as the reference.
At equilibrium, each state carries a relative statistical weight equal to its
formation-to-dissociation rate ratio, including the encounter weight, following
the same detailed-balance relation as in the closed ZEN system
(Section~\ref{closed}).
The relative weight of the state in which \(X_1\) binds \(X_2\) to form \(X_3\) is
\[
A = \frac{k_{\mathrm{evt}}^{(X_1+X_2\rightarrow X_3):S_1}\, w((X_1,X_2):S_1)}{\lambda_{\mathrm{evt}}^{(X_3\rightarrow X_1+X_2):S_1}},
\]
and that of the state in which \(X_1\) binds one \(X_4\) to form \(X_5\) is
\[
B = \frac{k_{\mathrm{evt}}^{(X_1+X_4\rightarrow X_5):S_1}\, w((X_1,X_4):S_1)}{\lambda_{\mathrm{evt}}^{(X_5\rightarrow X_1+X_4):S_1}}.
\]
With \(n\) interchangeable \(\widehat{X}_4\)-derived competitors, each contributing
the same relative weight \(B\), the unbound, \(X_3\), and \(X_5\) states carry
relative weights
\[
1, \qquad A, \qquad nB.
\]
The conditional probability that \(X_1\) is bound to \(X_2\) in \(S_1\) is therefore
\begin{equation}
P(N_{X3}(S_1)=1 \mid N_{\widehat{X4}}(S_1)=n) = \frac{A}{1+A+nB}.
\label{eq:cond_prob}
\end{equation}
Leaving the form of \(\pi(n)\) unspecified, the observed binding probability is
\begin{equation}
P(N_{X3}(S_1)=1) = \sum_n \pi(n)\, \frac{A}{1+A+nB}.
\label{eq:P_X3}
\end{equation}
Here \(\pi(n)\) is the distribution of the number of \(\widehat{X}_4\) moieties in
\(S_1\). The following subsections give \(\pi(n)\) for specific occupancy models.

\subsection{Derivation of the ${}_1F_1$ Form from a Poisson Occupancy} %
\label{subsec:poisson_1f1_derivation}

The previous subsection gave the conditional response at fixed
\(N_{\widehat{X4}}(S_1)=n\) and expressed the observed response as an average over
a general occupancy distribution \(\pi(n)\). Here we take \(\pi(n)\) to be Poisson.

This choice applies when the \(\widehat{X}_4\) moieties are sufficiently dilute,
enter \(S_1\) independently of one another, and face no explicit cap on their
number in \(S_1\). The occupancy \(N_{\widehat{X4}}(S_1)\) is then approximately
Poisson with mean \(\lambda\),
\begin{equation}
\pi_{\mathrm{P}}(n)=P(N_{\widehat{X4}}(S_1)=n)=e^{-\lambda}\frac{\lambda^n}{n!},
\qquad n=0,1,2,\ldots
\label{eq:pi_poisson}
\end{equation}
where \(\lambda\) is the mean occupancy of \(S_1\) across windows or cavities.

The conditional binding probability from the previous subsection,
\[
P(N_{X3}(S_1)=1\mid N_{\widehat{X4}}(S_1)=n)=\frac{A}{1+A+nB},
\]
can be rewritten as
\[
\frac{A}{1+A+nB}=\frac{A}{1+A}\,\frac{\nu}{\nu+n}, \qquad \nu=\frac{1+A}{B},
\]
so the occupancy-dependent normalized response is \(R(n)=\nu/(\nu+n)\); the shape
parameter \(\nu\) is the competitor count at which this response falls to half its
competitor-free value.

Averaging over the Poisson occupancy gives
\begin{align}
\bar{R}_{\mathrm{P}}
&=\sum_{n=0}^{\infty}\pi_{\mathrm{P}}(n)\frac{\nu}{\nu+n}
=e^{-\lambda}\sum_{n=0}^{\infty}\frac{\lambda^n}{n!}\frac{\nu}{\nu+n}.
\end{align}
Using \(\frac{\nu}{\nu+n}=\nu\int_0^1 t^{\nu+n-1}\,dt\),
\begin{align}
\bar{R}_{\mathrm{P}}
&=\nu e^{-\lambda}\int_0^1 t^{\nu-1}\sum_{n=0}^{\infty}\frac{(\lambda t)^n}{n!}\,dt
=\nu e^{-\lambda}\int_0^1 t^{\nu-1}e^{\lambda t}\,dt.
\end{align}
This integral is a confluent hypergeometric function,
\begin{equation}
\bar{R}_{\mathrm{P}}=e^{-\lambda}\,{}_1F_1(\nu;\nu+1;\lambda),
\end{equation}
and Kummer's transformation gives
\begin{equation}
\bar{R}_{\mathrm{P}}={}_1F_1(1;\nu+1;-\lambda).
\end{equation}
Restoring the prefactor \(A/(1+A)\), the binding probability is
\[
P(N_{X3}(S_1)=1)=\frac{A}{1+A}\,{}_1F_1(1;\nu+1;-\lambda).
\]

\subsection{Dose–Response Expression and Connection to Kd}
\label{subsec:Kd}

The previous subsection showed that, for Poisson occupancy,
\[
P(N_{X3}(S_1)=1)=\frac{A}{1+A}\,{}_1F_1(1;\nu+1;-\lambda), \qquad \nu=\frac{1+A}{B}.
\]
Here we rewrite this expression in terms of \(K_d\) and \(c_{\mathrm{eff}}\), making
its correspondence with the experimental dose--response expression explicit.

The mean occupancy \(\lambda\) introduced in Eq.~\eqref{eq:pi_poisson} follows
from the steady-state partitioning of \(\widehat{X}_4\) between the reservoir
\(S_0\) and the cavity \(S_1\).
Let \(\alpha_{X4:(S_0\rightleftarrows S_1)}\) be the dimensionless equilibrium
partition coefficient of \(\widehat{X}_4\) for \(S_0\rightleftarrows S_1\), i.e.\
the ratio of its \(S_1\) concentration to its \(S_0\) concentration. In a kinetic
exchange description, this coefficient can be related to the steady-state exchange
between \(S_0\) and \(S_1\), but here we use it as a measurable input rather than
deriving the exchange kinetics. The mean occupancy is then
\begin{equation}
\lambda
=
\alpha_{X4:(S_0\rightleftarrows S_1)}
\frac{c_{X4}(S_0)}
{c_{\mathrm{eff}}^{(X_1,X_4):S_1}}.
\label{eq:lambda_partition}
\end{equation}
Here, \(c_{X4}(S_0)\) denotes the molar concentration of free \(X_4\) in the
reservoir \(S_0\). Thus,
\(\alpha_{X4:(S_0\rightleftarrows S_1)}\,c_{X4}(S_0)\) is the local \(S_1\)
concentration associated with the conserved \(\widehat{X}_4\) moiety count.
Dividing this concentration by the single-competitor effective concentration
\(c_{\mathrm{eff}}^{(X_1,X_4):S_1}\) for the \((X_1,X_4)\) pair in \(S_1\)
(defined below in Eq.~\eqref{eq:ceff}) yields \(\lambda\), the mean number of
\(\widehat{X}_4\) moieties present in \(S_1\).

Specifically, analytical manipulation yields
\begin{equation}
\begin{aligned}
P(&N_{X3}(S_1)=1) =
\; \frac{\frac{k_{\mathrm{evt}}^{(X_1+X_2\rightarrow X_3):S_1}\,w((X_1,X_2):S_1)}{\lambda_{\mathrm{evt}}^{(X_3\rightarrow X_1+X_2):S_1}}}{1 + \frac{k_{\mathrm{evt}}^{(X_1+X_2\rightarrow X_3):S_1}\,w((X_1,X_2):S_1)}{\lambda_{\mathrm{evt}}^{(X_3\rightarrow X_1+X_2):S_1}}}\\[1mm]
&\quad \times {}_1F_1\!\Biggl(1;\;
\frac{1+\frac{k_{\mathrm{evt}}^{(X_1+X_2\rightarrow X_3):S_1}\,w((X_1,X_2):S_1)}{\lambda_{\mathrm{evt}}^{(X_3\rightarrow X_1+X_2):S_1}}}{\frac{k_{\mathrm{evt}}^{(X_1+X_4\rightarrow X_5):S_1}\,w((X_1,X_4):S_1)}{\lambda_{\mathrm{evt}}^{(X_5\rightarrow X_1+X_4):S_1}}}+1;\;
-\alpha_{X4:(S_0\rightleftarrows S_1)}\,\frac{c_{X4}(S_0)}{c_{\mathrm{eff}}^{(X_1,X_4):S_1}} \Biggr).
\end{aligned}
\label{eq:hypergeom}
\end{equation}
The third argument of \({}_1F_1\) corresponds to \(-\lambda\) in Eq.~\eqref{eq:lambda_partition}.

The effective concentrations are defined as
\begin{equation}
\begin{aligned}
c_{\mathrm{eff}}^{(X_1,X_2):S_1} &:= \frac{w((X_1,X_2):S_1)}{N_A\,\xi_{\mathrm{enc}}^{(X_1,X_2):S_1}}, \\
c_{\mathrm{eff}}^{(X_1,X_4):S_1} &:= \frac{w((X_1,X_4):S_1)}{N_A\,\xi_{\mathrm{enc}}^{(X_1,X_4):S_1}},
\end{aligned}
\label{eq:ceff}
\end{equation}
and the reaction equilibrium constants $K_d$ are expressed as
\begin{equation}
\frac{k_{\mathrm{evt}}^{(X_1+X_2\rightarrow X_3):S_1}\,w((X_1,X_2):S_1)}{\lambda_{\mathrm{evt}}^{(X_3\rightarrow X_1+X_2):S_1}}
=\frac{c_{\mathrm{eff}}^{(X_1,X_2):S_1}}{K_d^{X_1,X_2}},\quad
\frac{k_{\mathrm{evt}}^{(X_1+X_4\rightarrow X_5):S_1}\,w((X_1,X_4):S_1)}{\lambda_{\mathrm{evt}}^{(X_5\rightarrow X_1+X_4):S_1}}
=\frac{c_{\mathrm{eff}}^{(X_1,X_4):S_1}}{K_d^{X_1,X_4}}.
\label{eq:Kd}
\end{equation}
Substituting these into Eq.~(\ref{eq:hypergeom}) yields
\begin{equation}
\begin{aligned}
P(N_{X3}(S_1)=1) &=
\; \frac{\displaystyle \frac{c_{\mathrm{eff}}^{(X_1,X_2):S_1}}{K_d^{X_1,X_2}}}{\displaystyle 1+\frac{c_{\mathrm{eff}}^{(X_1,X_2):S_1}}{K_d^{X_1,X_2}}} {}_1F_1\!\Biggl(1;\;
\frac{1+\frac{c_{\mathrm{eff}}^{(X_1,X_2):S_1}}{K_d^{X_1,X_2}}}{\frac{c_{\mathrm{eff}}^{(X_1,X_4):S_1}}{K_d^{X_1,X_4}}}+1;\;
-\alpha_{X4:(S_0\rightleftarrows S_1)}\,\frac{c_{X4}(S_0)}{c_{\mathrm{eff}}^{(X_1,X_4):S_1}} \Biggr)\,.
\end{aligned}
\label{eq:final}
\end{equation}

Equation~(\ref{eq:final}) represents the final expression for the binding probability, maintaining the hypergeometric function representation while incorporating $K_d$ and $c_{\mathrm{eff}}$.
Note that when $X_4$ is assumed to be uniformly distributed within the ZEN system,
$
\int d^3\bm{x}\, f_{X1}(\bm{x};t) f_{X4}(\bm{x};t)=\frac{1}{V(S_1)}
$
holds, simplifying to $c_{\mathrm{eff}}^{(X_1,X_4):S_1}=\frac{1}{N_A V(S_1)}$.

When the occupancy distribution $\pi(n)$ is sharply concentrated around its mean, the sum is well approximated by the response evaluated at $\mathbb{E}_{\pi}[n]$, yielding a rational function of concentration. In the present system, copy numbers are of order one, so the full distribution contributes and the sum yields ${}_1F_1$. The rational expression is recovered as a limiting case.

\subsection{Derivation of the ${}_2F_1$ Form from a Binomial Occupancy} %
\label{subsec:binomial_2f1_derivation}

The previous subsection assumed Poisson occupancy, placing no explicit cap on
the number of \(\widehat{X}_4\) moieties in \(S_1\). Here we consider a finite
number of candidate moieties that can contribute to \(S_1\).

Let \(M\) be the number of candidates, each present in \(S_1\) independently with
probability \(p\). This corresponds to the approximation that the
\(\widehat{X}_4\) moieties are dilute enough to neglect correlations between
candidates, while the number that \(S_1\) can hold is finite. The occupancy
\(N_{\widehat{X4}}(S_1)\) is then approximately binomial,
\begin{equation}
\pi_{\mathrm{B}}(n)=P(N_{\widehat{X4}}(S_1)=n)={M \choose n}p^n(1-p)^{M-n},
\qquad n=0,1,\ldots,M
\label{eq:pi_binomial}
\end{equation}
with mean occupancy \(\mathbb{E}[N_{\widehat{X4}}(S_1)]=Mp\).

As before, the conditional binding probability is
\[
\frac{A}{1+A+nB}=\frac{A}{1+A}\,\frac{\nu}{\nu+n}, \qquad \nu=\frac{1+A}{B},
\]
so the normalized response at fixed \(N_{\widehat{X4}}(S_1)=n\) is
\begin{equation}
R(n)=\frac{\nu}{\nu+n}.
\end{equation}

Averaging over the binomial occupancy gives
\begin{align}
\bar{R}_{\mathrm{B}}
&=\sum_{n=0}^{M}\pi_{\mathrm{B}}(n)\frac{\nu}{\nu+n}
=\sum_{n=0}^{M}{M \choose n}p^n(1-p)^{M-n}\frac{\nu}{\nu+n}.
\end{align}
Again using \(\frac{\nu}{\nu+n}=\nu\int_0^1 t^{\nu+n-1}\,dt\),
\begin{align}
\bar{R}_{\mathrm{B}}
&=\nu\int_0^1 t^{\nu-1}\sum_{n=0}^{M}{M \choose n}(pt)^n(1-p)^{M-n}\,dt
=\nu\int_0^1 t^{\nu-1}(1-p+pt)^M\,dt.
\end{align}
In terms of the Gauss hypergeometric function,
\begin{equation}
\bar{R}_{\mathrm{B}}={}_2F_1(-M,1;\nu+1;p).
\end{equation}
Thus, when the local occupancy is set by a finite number of candidates, the
averaged inhibition response is a \({}_2F_1\) function. Restoring the prefactor
\(A/(1+A)\), as in the Poisson case, gives the binding probability
\[
P(N_{X3}(S_1)=1)=\frac{A}{1+A}\,{}_2F_1(-M,1;\nu+1;p).
\]

In the limit \(M\to\infty\), \(p\to0\) with \(Mp=\lambda\) fixed, the binomial
distribution converges to the Poisson distribution, and
\begin{equation}
{}_2F_1(-M,1;\nu+1;p)\to{}_1F_1(1;\nu+1;-\lambda),
\end{equation}
recovering the Poisson result.

\subsection{Rational versus Hypergeometric Form} %
\label{subsec:rational_vs_hypergeometric}

The preceding subsections showed that, when the occupancy distribution \(\pi(n)\)
is sharply concentrated about its mean, the response is well approximated by the
rational form evaluated at \(\mathbb{E}_{\pi}[n]\); when the copy number is of
order one and the whole distribution contributes, as here, a hypergeometric form
arises instead. This hypergeometric form corresponds to Eqs.~\eqref{eq:hypergeom}
and \eqref{eq:final}, and the underlying observable is the conditional binding
probability Eq.~\eqref{eq:cond_prob} averaged over \(\pi(n)\), namely
Eq.~\eqref{eq:P_X3}. We now compare the two forms: which approximation the
rational form represents, and why the present system needs the hypergeometric
form.

The difference reflects at which stage the local \(X_4\) count is averaged. In the
open ZEN system, \(X_4\) exchanges with the exterior, so the conserved count
\(N_{\widehat{X4}}(S_1)\) in stage \(S_1\) is a random variable with the occupancy
distribution \(\pi(n)\) (Eqs.~\eqref{eq:pi_poisson} and \eqref{eq:pi_binomial}).
The observable is obtained by first computing the conditional binding probability
at fixed \(N_{\widehat{X4}}(S_1)=n\) and then averaging over \(\pi(n)\),
\begin{equation}
P(N_{X3}(S_1)=1)=\mathbb{E}_{\pi}\!\left[P(N_{X3}(S_1)=1\mid N_{\widehat{X4}}(S_1)=n)\right]
\end{equation}
(Eq.~\eqref{eq:P_X3}). What matters is therefore not the mean count itself, but
evaluating the response while keeping the local count distribution \(\pi(n)\).

The rational form, by contrast, replaces \(n\) in Eq.~\eqref{eq:cond_prob} by its
mean. Taking \(\mathbb{E}_{\pi}[n]\) and treating every stage as carrying the same
mean occupancy gives
\begin{equation}
\begin{aligned}
P(N_{X3}(S_1)=1)\approx
\frac{\displaystyle \frac{k_{\mathrm{evt}}^{(X_1+X_2\rightarrow X_3):S_1}\,w((X_1,X_2):S_1)}{\lambda_{\mathrm{evt}}^{(X_3\rightarrow X_1+X_2):S_1}}}
{\displaystyle 1+\frac{k_{\mathrm{evt}}^{(X_1+X_2\rightarrow X_3):S_1}\,w((X_1,X_2):S_1)}{\lambda_{\mathrm{evt}}^{(X_3\rightarrow X_1+X_2):S_1}}
+\mathbb{E}_{\pi}[n]
\frac{k_{\mathrm{evt}}^{(X_1+X_4\rightarrow X_5):S_1}\,w((X_1,X_4):S_1)}{\lambda_{\mathrm{evt}}^{(X_5\rightarrow X_1+X_4):S_1}}}\,,
\end{aligned}
\end{equation}
and Eq.~\eqref{eq:Kd} recasts the coefficients into the \(c_{\mathrm{eff}}/K_d\)
form. This approximation holds when the local count distribution is sharp and
every compartment carries nearly the same occupancy.

When the local copy number fluctuates at order one, however, the right-hand side
of Eq.~\eqref{eq:cond_prob} is nonlinear in \(n\), so averaging and substitution
cannot be interchanged:
\begin{equation}
\begin{aligned}
\mathbb{E}_{\pi}\!\Biggl[
&\frac{\displaystyle \frac{k_{\mathrm{evt}}^{(X_1+X_2\rightarrow X_3):S_1}\,w((X_1,X_2):S_1)}{\lambda_{\mathrm{evt}}^{(X_3\rightarrow X_1+X_2):S_1}}}
{\displaystyle 1+\frac{k_{\mathrm{evt}}^{(X_1+X_2\rightarrow X_3):S_1}\,w((X_1,X_2):S_1)}{\lambda_{\mathrm{evt}}^{(X_3\rightarrow X_1+X_2):S_1}}
+n\,\frac{k_{\mathrm{evt}}^{(X_1+X_4\rightarrow X_5):S_1}\,w((X_1,X_4):S_1)}{\lambda_{\mathrm{evt}}^{(X_5\rightarrow X_1+X_4):S_1}}}
\Biggr] \\
&\neq
\frac{\displaystyle \frac{k_{\mathrm{evt}}^{(X_1+X_2\rightarrow X_3):S_1}\,w((X_1,X_2):S_1)}{\lambda_{\mathrm{evt}}^{(X_3\rightarrow X_1+X_2):S_1}}}
{\displaystyle 1+\frac{k_{\mathrm{evt}}^{(X_1+X_2\rightarrow X_3):S_1}\,w((X_1,X_2):S_1)}{\lambda_{\mathrm{evt}}^{(X_3\rightarrow X_1+X_2):S_1}}
+\mathbb{E}_{\pi}[n]
\frac{k_{\mathrm{evt}}^{(X_1+X_4\rightarrow X_5):S_1}\,w((X_1,X_4):S_1)}{\lambda_{\mathrm{evt}}^{(X_5\rightarrow X_1+X_4):S_1}}}\,.
\end{aligned}
\end{equation}
This non-commutativity is the direct origin of the difference between the two
forms. The hypergeometric form derived here resums the weighted series while
preserving the distribution average, and corresponds to Eqs.~\eqref{eq:hypergeom}
and \eqref{eq:final}.

Which of the two forms applies therefore depends on whether the occupancy
distribution \(\pi(n)\) is adequately represented by its mean
\(\mathbb{E}_{\pi}[n]\). In the present system, where the local copy number is of
order one, the full distribution contributes and the hypergeometric form is
required.

\section{Bayesian response-model fitting}
\label{sec:bayesian_kd_smbit_lgbit}

The hypergeometric response functions derived above were used as Bayesian
response models for the SmBiT--LgBiT/DarkBiT titration data in Fig.~5. This
section describes how the \({}_1F_1\) and finite-capacity \({}_2F_1\) kernels
were connected to the titration curves, how shared and series-specific
parameters were estimated, and how sampling convergence and predictive
performance were assessed. The external affinity calibration and numerical
SmBiT230 \(K_{\mathrm d}\) estimate are reported in the Supplementary
Information of the companion experimental study\cite{CompanionPaper}.

The fit shown in Fig.~5a used the \({}_1F_1\) response kernel and a two-stage
fitting procedure. The fit shown in Fig.~5c used the finite-capacity
\({}_2F_1\) response kernel and the external-affinity calibration specified in
the companion experimental SI. Both panels therefore use the response models
defined below, whereas the affinity-calibration details are kept with the
corresponding \(K_{\mathrm d}\) analysis.

\subsection{Empirical parameterization of the titration response}
\label{subsec:parameter_source_smbit}

For each SmBiT series \(i\), the DarkBiT titration concentration is denoted by
\(c_{ij}\), the normalized observed luminescence by \(y_{ij}\), and the model
mean by
\begin{equation}
  \mu_{ij}=s_i P_i(c_{ij}),
  \label{eq:mean}
\end{equation}
where \(s_i\) is a series-specific signal scale and \(P_i(c)\) is the response
function. Replicate-level normalized values were used in the likelihood. Each
replicate was treated as conditionally independent given the model parameters.
Residual variation associated with normalization was represented by the
Student-\(t\) observation scales defined below.

The SmBiT--LgBiT dissociation constant enters through the zero-DarkBiT
occupancy parameter
\begin{equation}
  \rho_i=\frac{c_{\mathrm{eff}}}{K_{\mathrm d,i}},
  \label{eq:rho}
\end{equation}
so that
\begin{equation}
  P_i(0)=\frac{\rho_i}{1+\rho_i}.
\end{equation}
Here \(c_{\mathrm{eff}}\) is the shared SmBiT--LgBiT effective-concentration
scale used to convert an affinity into a dimensionless local occupancy. It is
a response-model parameter and was estimated jointly in the finite-capacity
fit. Its external calibration role, and the distinction from the
DarkBiT--LgBiT geometric scale estimated from the spatial-overlap calculation
underlying Fig.~4b, are described in the companion experimental SI.

\subsection{Hypergeometric response functions}

The \({}_1F_1\) and finite-capacity \({}_2F_1\) forms arise by averaging the
conditional binding response over Poisson and binomial local competitor
occupancies, respectively, as derived in
Sections~\ref{subsec:poisson_1f1_derivation} and
\ref{subsec:binomial_2f1_derivation}. They are rewritten here in the
parameterization used for fitting.

For the Poisson-occupancy model, the target-state weight \(A\) was identified
with \(\rho_i=c_{\mathrm{eff}}/K_{\mathrm d,i}\), the mean competitor
occupancy was written as \(\lambda(c)=\phi c\), and
\(\nu_i=(1+\rho_i)\psi_i\). The resulting response kernel was
\begin{equation}
  P_i^{(1F1)}(c)
  =
  \frac{\rho_i}{1+\rho_i}
  {}_1F_1\!\left(1;\nu_i+1;-\phi c\right),
  \qquad
  \nu_i=(1+\rho_i)\psi_i .
  \label{eq:1f1}
\end{equation}
For the \({}_1F_1\) fit, both \(\psi_i=\psi\) and \(\phi\) were shared across
series.

For the finite-capacity model, the binomial occupancy probability was
parameterized as
\begin{equation}
  p(c)=\frac{\phi c}{1+\phi c},
  \label{eq:binomial_occupancy_link}
\end{equation}
so that the mean local occupancy is \(Mp(c)\). The response kernel was
\begin{equation}
  P_i^{(2F1)}(c)
  =
  \frac{\rho_i}{1+\rho_i}
  {}_2F_1\!\left(
    -M,1;\nu_i+1;\frac{\phi c}{1+\phi c}
  \right),
  \qquad
  \nu_i=(1+\rho_i)\psi_i .
  \label{eq:2f1}
\end{equation}
The reported fit used \(M=24\), selected after exploratory comparisons among
several candidate integer values. Because \(M\) was fixed rather than sampled,
its selection uncertainty is not included in the posterior intervals. The
negative integer argument \(-M\) makes the hypergeometric function a finite
polynomial.

For the \({}_2F_1\) fit, the series-specific shape parameter was written as
\begin{equation}
  \log_{10}\nu_i
  =
  \log_{10}\nu_0
  +
  \beta_\nu
  \left[
    \log_{10}\left(\frac{K_{\mathrm d,i}}{\mathrm M}\right)-x_0
  \right],
  \qquad
  \psi_i=\frac{\nu_i}{1+\rho_i},
  \label{eq:nu-beta}
\end{equation}
with \(\beta_\nu=-0.5\), \(x_0=-6\), and a shared \(\phi\).
The coefficient \(\beta_\nu\) was fixed after exploratory model checks to
stabilize the affinity-dependent curve-shape trend. Consequently, the
\({}_2F_1\) fit is conditional on \(M\), \(\beta_\nu\), \(x_0\), and the
shared-\(\phi\) parameterization.

\subsection{Bayesian observation model}

The replicate-level likelihood was
\begin{equation}
  y_{ij}
  \sim
  \operatorname{Student}t_4\!\left(\mu_{ij},\sigma_{d(i)}\right),
  \label{eq:likelihood}
\end{equation}
where \(d(i)\) denotes the dataset class. The \({}_1F_1\) analysis included
primary and linker-length titrations and therefore fitted separate
\(\sigma_{\mathrm{primary}}\) and \(\sigma_{\mathrm{linker}}\) values. The
\({}_2F_1\) analysis used the six primary titration series and fitted only
\(\sigma_{\mathrm{primary}}\). The Student-\(t_4\) distribution limits the
influence of occasional replicate-level deviations.

The primary finite-capacity fit used a shape-weighted objective,
\begin{equation}
  \log\widetilde p(D\mid\theta)
  =
  \sum_{ij}\omega_{ij}\log p(y_{ij}\mid\theta),
  \label{eq:weighted_objective}
\end{equation}
with the highest and lowest 25\% of the positive concentration levels in each
series defining the tail and head groups. Every observation had a base weight
of one. For each series, an additional total weight of 10 was divided equally
among the tail observations and a total weight of 0.5 among the head
observations; the zero-concentration point was excluded from the head group.
This objective was used to retain the high-concentration curve shapes in the
simultaneous fit. It defines a weighted
pseudo-posterior, not an additional generative observation model; the
resulting intervals are therefore weighted pseudo-posterior intervals rather
than nominal credible intervals for the unweighted Student-\(t_4\) model.
Equal-weight and alternative mean-one weighting results are reported as
sensitivity analyses in the companion experimental SI.

\subsection{Priors and fixed values}

Concentrations were expressed in molar units and \(\phi\) in
\(\mathrm{M}^{-1}\) when evaluating the log-scale priors. The fitting
parameters were assigned the following priors, where applicable:
\begin{align}
  \log_{10}\left(\frac{c_{\mathrm{eff}}}{\mathrm M}\right)
  &\sim
  \mathcal{N}(-5,1.5^2)\ \mathrm{truncated\ to}\ [-12,2], \\
  \log_{10}\left(\frac{\phi}{\mathrm M^{-1}}\right)
  &\sim
  \mathcal{N}(5.2,0.5^2)\ \mathrm{truncated\ to}\ [4,8.5], \\
  \log_{10}\psi
  &\sim
  \mathcal{N}(-1.5,0.4^2)\ \mathrm{truncated\ to}\ [-4,0.5],
  &&\text{for the shared-\(\psi\) \({}_1F_1\) fit}, \\
  \log_{10}\nu_0
  &\sim
  \mathcal{N}(-0.8,0.6^2)\ \mathrm{truncated\ to}\ [-6,3],
  &&\text{for the \({}_2F_1\) fit}, \\
  \log_{10}s_i
  &\sim
  \mathcal{N}(\log_{10}s_{i,\mathrm{anchor}},0.2^2), \\
  \log_{10}\rho_i
  &\sim
  \mathcal{N}(0,1.5^2)\ \mathrm{truncated\ to}\ [-4,6],
  &&\text{for free \(\rho_i\) in the \({}_1F_1\) second stage}, \\
  \log_{10}\left(\frac{K_{\mathrm d,i}}{\mathrm M}\right)
  &\sim
  \operatorname{Uniform}(-12,-1),
  &&\text{for free affinities in the \({}_2F_1\) fit}, \\
  \sigma_d
  &\sim
  \operatorname{Exponential}(0.05)\
  \mathrm{truncated\ to}\ [10^{-6},10^2].
\end{align}
Here \(s_{i,\mathrm{anchor}}\) is the configured reference signal scale used
to centre the weakly informative prior on \(s_i\). The broad truncation limits
were numerical regularizers; the reported posterior distributions were not
concentrated at these limits. SmBiT114 and SmBiT230 signal scales were fixed
at one in the finite-capacity fit. The affinity anchors and
external-observation model used to calibrate the \({}_2F_1\) fit are specified
in the companion experimental SI.

\subsection{Two-stage \texorpdfstring{\({}_1F_1\)}{1F1} fit for Figure 5a}

Stage 1 inferred \(c_{\mathrm{eff}}\) from the primary SmBiT99, SmBiT104, and
SmBiT114 curves using their reference \(K_{\mathrm d}\) values. The posterior
median,
\(c_{\mathrm{eff}}=7.18\times10^{-4}\,\mathrm M\)
\([6.44\times10^{-4},8.03\times10^{-4}\,\mathrm M]\), was held fixed in
stage 2.

The second-stage fit used the primary SmBiT99, SmBiT104, SmBiT114, and
SmBiT230 titration curves together with four SmBiT114 linker-length titration
curves. It fitted shared \(\psi\) and \(\phi\), series-specific signal scales,
and free \(\rho\) values for SmBiT230 and the non-reference linker-length
conditions. The
\(4.1\,\mathrm{nm}\)-LgBiT/\(4.1\,\mathrm{nm}\)-SmBiT114 condition shared
the SmBiT114 occupancy reference.

\subsection{Finite-capacity \texorpdfstring{\({}_2F_1\)}{2F1} fit for Figure 5c}

The fit used the six primary SmBiT titration series: SmBiT86, SmBiT78,
SmBiT99, SmBiT104, SmBiT114, and SmBiT230. The linker-length series were not
included. The shared effective concentration, the four freely estimated affinities, and
the remaining response parameters were estimated simultaneously under the
external-observation calibration model. The calibration assumptions and
numerical affinity results are given in the companion experimental SI; the
present fit supplies the finite-capacity response curves shown in Fig.~5c.

\subsection{Sampling diagnostics and posterior predictive checks}

The \({}_1F_1\) fit used the No-U-Turn Sampler (NUTS) in Turing.jl with four
chains, 2,000 warmup iterations, and 4,000 retained samples per chain. Its
maximum monitored \(\widehat R\) was 1.0006, its minimum bulk and tail
effective sample sizes were 6,634 and 6,148, respectively, and no divergent
transitions were observed.

The shape-weighted \({}_2F_1\) fit used four chains, 2,000 warmup iterations,
4,000 retained samples per chain, and a target acceptance probability of
0.95. Its maximum monitored \(\widehat R\) was 1.0013, its minimum bulk and
tail effective sample sizes were 4,800 and 3,704, respectively, and no
divergent transitions were observed.

Posterior mean-response bands and posterior predictive bands were generated
for both fits at the 95\% level. These checks were used to compare the fitted curve shapes with
the replicate-level titration observations. The posterior predictive bands
covered most replicate-level observations; the remaining outliers were
concentrated in a small number of series--concentration combinations rather
than indicating a global lack of fit.

\subsection{Reproducibility}

The analyses were implemented in Julia and Turing.jl. The response-model
implementation and figure-redraw scripts are provided with the theory-paper
code, while the affinity-calibration configurations, numerical summaries, and
calibration sensitivity analyses are provided with the experimental-paper
code.

\section{Entropic interpretation}
\label{ap_entropy}
\subsection{Equilibrium analysis through differential entropy}

The preceding sections obtained the binding equilibria kinetically, from the
weighted-graph rate balance. Here we give an independent thermodynamic derivation
of the same equilibria and show that the encounter weight \(w\) carries a
definite entropic meaning.

For an entity probability distribution \(f(x)\), the
differential entropy is:
\begin{equation}
h[f] := - \int f(x) \log f(x) dx.
\end{equation}
Differential entropy connects to thermodynamic entropy through:
\begin{equation}
S_{\mathrm{position}}[f] := k_B (h[f] - \log \xi).
\label{S-h}
\end{equation}
Because differential entropy depends on the choice of reference units for
continuous variables, \(\xi\) supplies the reference measure. Only the combined
quantity \(h[f]-\log \xi\) is physically meaningful as a positional entropy
contribution. In the uniform spatial case this is the logarithm of a
dimensionless measure ratio, \(\log(V/\xi)\).
The total entropy incorporates both positional and internal contributions:
\begin{equation}
S = S_{\mathrm{internal}} + S_{\mathrm{position}}[f] = S_{\mathrm{internal}} + k_B (h[f] - \log \xi).
\end{equation}
Here \(\xi\) is a correction term with volume dimensions (\([L^d]\) for
\(d\)-dimensional degrees of freedom) related to counting units; in
thermodynamic entropy, it plays the role of a reference cell size for continuous
degrees of freedom. The same construction applies in phase space. The ideal-gas
example below uses the full phase-space reference measure, but only its
positional projection enters \(S_{\mathrm{position}}[f]\). For an ideal gas:
\begin{equation}
\begin{aligned}
H^{ID} =& H^D(\bm{R}^N) + H^D(\bm{p}^N) - I_{\mathrm{qm}} - I_{\mathrm{indist}} \\
=& N\log V + \frac{3N}{2}\log (2\pi emT) - 3N \log h - \log N! \\
=& N\log \left[ \frac{V}{N}\left( \frac{2\pi mT}{h^2}\right) ^{3/2} \right] + \frac{5N}{2} ,
\end{aligned}
\label{free_gas}
\end{equation}
as established in Ref.~\cite{Ben2008}. Here \(I_{\mathrm{qm}}\) and
\(I_{\mathrm{indist}}\) denote the quantum-mechanical counting and
indistinguishability corrections, respectively.
The quantum-mechanical counting correction in Eq.~(\ref{free_gas}) is
\begin{equation}
I_{\mathrm{qm}} = 3N \log h = \log h^{3N},
\end{equation}
and the indistinguishability correction is
\begin{equation}
I_{\mathrm{indist}} = \log N!.
\end{equation}
The former corresponds to the phase-space version of the \(\log \xi\)
reference-cell term in Eq.~(\ref{S-h}).

The reaction entropy $\Delta S$ follows directly from these thermodynamic relations. For the closed ZEN system examined in section \ref{closed}, the reaction $X_1 + X_2 \ce{->} X_3$ yields:
\begin{equation}
\begin{aligned}
\Delta S :=& S_{\mathrm{products}} - S_{\mathrm{reactants}} \\
=& \Delta S_{\mathrm{internal}} + k_B((h[f_{X3}]-\log \xi) - (h[f_{X1}]-\log \xi) - (h[f_{X2}]-\log \xi)) \\
=& \Delta S_{\mathrm{internal}} + k_B(h[f_{X3}] - h[f_{X1}] - h[f_{X2}] + \log \xi),
\end{aligned}
\label{reaction_entropy}
\end{equation}
where
\begin{equation}
\Delta S_{\mathrm{internal}} := S_{\mathrm{internal}}^{X3} - (S_{\mathrm{internal}}^{X1} + S_{\mathrm{internal}}^{X2}).
\end{equation}

For uniform distributions $f_{\text{X1}}, f_{\text{X2}}, f_{\text{X3}}$ over volumes $V_{\text{X1}}, V_{\text{X2}}, V_{\text{X3}}$ respectively, using the differential entropy $\log V$ of a uniform distribution over volume $V$:
\begin{equation}
\Delta S = \Delta S_{\mathrm{internal}} + k_B \log \frac{\xi V_{\text{X3}}}{V_{\text{X1}} V_{\text{X2}}}.
\end{equation}

Since the positional support of the complex is confined to the accessible region of its constituents (i.e., $\mathrm{supp}(f_{\text{X3}}) \subseteq \mathrm{supp}(f_{\text{X1}}) \cap \mathrm{supp}(f_{\text{X2}})$), the weight $w(X_1,X_2)$ becomes:
\begin{equation}
\begin{aligned}
w(X_1, X_2) =& \xi_{\mathrm{enc}} \int_{\mathrm{supp}(f_{\text{X3}})} \frac{1}{V_{\text{X1}}}\cdot \frac{1}{V_{\text{X2}}} dx \\
=& \frac{\xi_{\mathrm{enc}} V_{\text{X3}}}{V_{\text{X1}} V_{\text{X2}}}.
\end{aligned}
\end{equation}
We now choose the thermodynamic reference cell \(\xi\) to match the encounter
coarse-graining scale \(\xi_{\mathrm{enc}}\). This identifies the reference
measure used in the entropy expression with the scale used in the encounter
weight, rather than introducing an additional physical assumption. Setting
\(\xi=\xi_{\mathrm{enc}}\):
\begin{equation}
\Delta S = \Delta S_{\mathrm{internal}} + k_B \log w(X_1, X_2).
\end{equation}

Using notation from equation (\ref{particle_state}):
\begin{equation}
\begin{aligned}
p := \pi(\bm{n}_1) \\
1-p = \pi(\bm{n}_0).
\end{aligned}
\end{equation}
The total system entropy uses per-entity entropies $S_{\bm{n}0},S_{\bm{n}1}$ for entity number states $\bm{n}_0, \bm{n}_1$:
\begin{equation}
S = k_B H_2(p) + (1-p) S_{\bm{n}0} + p S_{\bm{n}1},
\end{equation}
where $H_2(p):=-p \log p - (1-p) \log (1-p)$ denotes the two-state Shannon entropy.

The Helmholtz free energy $F$ incorporates per-entity energies $E_{\bm{n}0},E_{\bm{n}1}$:
\begin{equation}
F = (1-p) E_{\bm{n}0} + p E_{\bm{n}1} - TS.
\end{equation}
Here $E_{\bm{n}0}$ and $E_{\bm{n}1}$ denote the coarse-grained energies
assigned to the unbound and bound entity-number states, respectively, and
\begin{equation}
\Delta E := E_{\bm{n}1}-E_{\bm{n}0}.
\end{equation}
The reaction entropy defined above is the entropy difference between the
bound and unbound entity-number states:
\begin{equation}
\Delta S = S_{\bm{n}1}-S_{\bm{n}0}.
\end{equation}
For the uniform case considered above,
\begin{equation}
\Delta S
=
\Delta S_{\mathrm{internal}}
+
k_B \log w(X_1,X_2).
\end{equation}
At equilibrium:
\begin{equation}
\begin{aligned}
0=\frac{dF}{dp}
&=
\Delta E
-
k_B T \log \frac{1-p}{p}
-
T\Delta S \\
&=
\Delta E
-
T\Delta S_{\mathrm{internal}}
-
k_B T \log \frac{1-p}{p}
-
k_B T \log w(X_1,X_2).
\end{aligned}
\end{equation}
Therefore,
\begin{equation}
\frac{p}{1-p}
=
\exp\left(
-\frac{\Delta E}{k_B T}
+
\frac{\Delta S_{\mathrm{internal}}}{k_B}
\right)
w(X_1,X_2).
\end{equation}
Using the detailed-balance identification
\begin{equation}
\frac{k_{\mathrm{evt}}^{X1+X2\rightarrow X3}}
{\lambda_{\mathrm{evt}}^{X3\rightarrow X1+X2}}
=
\exp\left(
-\frac{\Delta E}{k_B T}
+
\frac{\Delta S_{\mathrm{internal}}}{k_B}
\right),
\label{evt_ratio_thermo_uniform}
\end{equation}
this yields:
\begin{equation}
p = \pi(\bm{n}_1) =
\frac{\displaystyle
\frac{k_{\mathrm{evt}}^{X1+X2\rightarrow X3}\, w(X_1,X_2)}
{\lambda_{\mathrm{evt}}^{X3\rightarrow X1+X2}}}
{\displaystyle
1+
\frac{k_{\mathrm{evt}}^{X1+X2\rightarrow X3}\, w(X_1,X_2)}
{\lambda_{\mathrm{evt}}^{X3\rightarrow X1+X2}}}.
\label{kwxl_michaelis}
\end{equation}
With the identification
$c_{\mathrm{eff}} := w(X_1,X_2)/(N_A\,\xi_{\mathrm{enc}}^{X1,X2})$,
this expression can be written in the standard Michaelis--Menten form
$p = c_{\mathrm{eff}}/(K_d+c_{\mathrm{eff}})$, where
\begin{equation}
K_d :=
\frac{\lambda_{\mathrm{evt}}^{X3\rightarrow X1+X2}}
{N_A\,\xi_{\mathrm{enc}}^{X1,X2}
 k_{\mathrm{evt}}^{X1+X2\rightarrow X3}}.
\end{equation}

\subsection{Inhomogeneous case}
The preceding section confined discussion to uniform probability distributions.
This section extends the analysis to inhomogeneous distributions by examining local probability flux balance.

\subsubsection{Local probability flux balance}

Consider a small coarse-graining cell of volume $\xi$ surrounding point $x$.
Because the forward reaction is bimolecular whereas the reverse reaction is unimolecular, the corresponding local reaction fluxes scale as $\xi^2$ and $\xi$, respectively.
The forward reaction rate $X_1 + X_2 \ce{->} X_3$ within this region takes the form
\begin{equation}
k_{\mathrm{evt}} f_{X1}(x) f_{X2}(x) \pi(\bm{n}_0)\, \xi^2,
\end{equation}
while the reverse reaction rate $X_3 \ce{->} X_1 + X_2$ is
\begin{equation}
\lambda_{\mathrm{evt}} f_{X3}(x) \pi(\bm{n}_1)\, \xi.
\end{equation}
At equilibrium, these local reaction fluxes balance:
\begin{equation}
\begin{aligned}
k_{\mathrm{evt}} f_{X1}(x) f_{X2}(x) \pi(\bm{n}_0)\, \xi^2
&= \lambda_{\mathrm{evt}} f_{X3}(x) \pi(\bm{n}_1)\, \xi \\
\iff\quad
f_{X3}(x)
&=
\frac{k_{\mathrm{evt}}\xi}{\lambda_{\mathrm{evt}}}
\frac{\pi(\bm{n}_0)}{\pi(\bm{n}_1)}
f_{X1}(x) f_{X2}(x).
\end{aligned}
\label{fX3}
\end{equation}
We identify the remaining coarse-graining factor \(\xi\) with the encounter
scale \(\xi_{\mathrm{enc}}\) below, so that it is absorbed into
\(w(X_1,X_2)\) rather than introduced as an independent scale.
The normalization condition for \(f_{X3}(x)\) requires
\begin{equation}
\int f_{X3}(x) dx = 1.
\end{equation}
Combining this with
\begin{equation}
w(X_1, X_2)
=
\xi_{\mathrm{enc}}
\int f_{X1}(x) f_{X2}(x)\,dx,
\label{wX1X2}
\end{equation}
and setting $\xi=\xi_{\mathrm{enc}}$, we obtain
\begin{equation}
\frac{k_{\mathrm{evt}} w(X_1, X_2)}
{\lambda_{\mathrm{evt}}}
\cdot
\frac{\pi(\bm{n}_0)}{\pi(\bm{n}_1)}
=
1.
\label{kwxlpp}
\end{equation}
Rearrangement of this expression recovers equation (\ref{kwxl_michaelis}), thus providing an alternative derivation of the Michaelis-Menten equation.

\subsubsection{Inhomogeneous entropy}
From equations (\ref{fX3}), (\ref{wX1X2}), and (\ref{kwxlpp}):
\begin{equation}
f_{X3}(x) = \frac{f_{X1}(x) f_{X2}(x)}{\displaystyle \int f_{X1}(x) f_{X2}(x) dx}.
\end{equation}
Substituting this expression into the reaction entropy equation
(\ref{reaction_entropy}) and using Eq.~\eqref{wX1X2} to rewrite the last logarithm:
\begin{equation}
\begin{aligned}
\Delta S =& \Delta S_{\mathrm{internal}} + k_B \Bigg( - \int dx \, f_{X3}(x) \log f_{X3}(x) + \int dx \, f_{X1}(x) \log f_{X1}(x) \\
&+ \int dx \, f_{X2}(x) \log f_{X2}(x) + \log \xi \Bigg) \\
=& \Delta S_{\mathrm{internal}} + k_B \Bigg( - \int dx \, f_{X3}(x) \log f_{X1}(x) - \int dx \, f_{X3}(x) \log f_{X2}(x) \\
&+ \int dx \, f_{X3}(x) \log \bigg( \int dx' \, f_{X1}(x') f_{X2}(x') \bigg) + \int dx \, f_{X1}(x) \log f_{X1}(x) \\
&+ \int dx \, f_{X2}(x) \log f_{X2}(x) + \log \xi \Bigg) \\
=& \Delta S_{\mathrm{internal}} + k_B \Bigg( \mathbb{E}_{f_{X3}} [-\log f_{X1}(x)] - \mathbb{E}_{f_{X1}} [-\log f_{X1}(x)] + \mathbb{E}_{f_{X3}} [-\log f_{X2}(x)] - \mathbb{E}_{f_{X2}} [-\log f_{X2}(x)] \\
&+ \log \left( \xi \int dx \, f_{X1}(x) f_{X2}(x) \right) \Bigg) \\
=& \Delta S_{\mathrm{internal}} + k_B (\mathbb{E}_{f_{X3}} [-\log f_{X1}(x)] - h[f_{X1}] + \mathbb{E}_{f_{X3}} [-\log f_{X2}(x)] - h[f_{X2}] + \log w(X_1, X_2)).
\end{aligned}
\end{equation}
The terms beyond $\Delta S_{\mathrm{internal}}$ and $k_B \log w(X_1, X_2)$ vanish under homogeneous conditions. We therefore call this additional contribution the inhomogeneous entropy, $\Delta S_{\mathrm{inh}}$:
\begin{equation}
\Delta S_{\mathrm{inh}}
= k_B \big(\mathbb{E}_{f_{X3}} [-\log f_{X1}(x)] - h[f_{X1}]
+ \mathbb{E}_{f_{X3}} [-\log f_{X2}(x)] - h[f_{X2}] \big).
\end{equation}
Thus:
\begin{equation}
\Delta S = \Delta S_{\mathrm{internal}} + \Delta S_{\mathrm{inh}} + k_B \log w(X_1, X_2).
\label{delta_s}
\end{equation}

To determine $\pi(\bm{n}_0)$ and $\pi(\bm{n}_1)$ at thermal equilibrium, we require the free energy to be stationary with respect to $p$:
\begin{equation}
\begin{aligned}
& \frac{dF}{dp}=0 \\
&\iff -k_B T \log \frac{1-p}{p} + \Delta E - T (\Delta S_{\mathrm{internal}} + \Delta S_{\mathrm{inh}} + k_B \log w(X_1, X_2)) = 0 \\
&\iff \frac{p}{1-p} = \exp \left( -\frac{\Delta E}{k_B T} + \frac{\Delta S_{\mathrm{internal}} + \Delta S_{\mathrm{inh}}}{k_B} \right) w(X_1, X_2).
\end{aligned}
\end{equation}
Using the relation
\begin{equation}
\frac{p}{1-p} = \frac{k_{\mathrm{evt}} w(X_1, X_2)}{\lambda_{\mathrm{evt}}},
\end{equation}
yields:
\begin{equation}
\frac{k_{\mathrm{evt}}}{\lambda_{\mathrm{evt}}} = \exp \left( -\frac{\Delta E}{k_B T} + \frac{\Delta S_{\mathrm{internal}} + \Delta S_{\mathrm{inh}}}{k_B} \right).
\end{equation}

\subsubsection{Scope of the independence approximation}

The analysis thus far assumes statistical independence between entities.
A rigorous treatment requires the joint probability distribution function
$f_{(1,i),(2,j)}(\bm{x}_1,\bm{x}_2;t)$ as defined in Eq.~(\ref{def_weight}).
Under the statistical-independence approximation $f_{(1,i),(2,j)}(\bm{x}_1,\bm{x}_2;t)\simeq f_{1,i}(\bm{x}_1;t)f_{2,j}(\bm{x}_2;t)$,
the weight reduces to (cf.\ Eq.~(\ref{def_weight_indep})):
\begin{equation}
\begin{aligned}
w(X_{1,i}, X_{2,j})
=\int d^3 \bm{x}_1 d^3 \bm{x}_2 \;
f_{1,i}(\bm{x}_1;t)\, f_{2,j}(\bm{x}_2;t)\,
\mathrm{Enc}(\mathrm{encounter};\bm{x}_1,\bm{x}_2,t).
\end{aligned}
\end{equation}
The general case requires:
\begin{equation}
\begin{aligned}
w(X_{1,i}, X_{2,j})
=\int d^3 \bm{x}_1 d^3 \bm{x}_2 \;
f_{(1,i),(2,j)}(\bm{x}_1,\bm{x}_2;t)\,
\mathrm{Enc}(\mathrm{encounter};\bm{x}_1,\bm{x}_2,t).
\end{aligned}
\end{equation}

More generally, correlations may exist not only between the reacting pair but also with additional entities in the system, necessitating correction terms derived from many-body joint distribution functions. Formally, such corrections can be organized using cluster-expansion techniques; keeping terms up to three-body correlations yields representative correction contributions of the form:
\begin{equation}
\begin{aligned}
v =& \sum_{i=1}^{N_{X1}} \sum_{j=1}^{N_{X2}} k_{\mathrm{evt}} \int d^3 \bm{x}_1 d^3 \bm{x}_2 \; f_{(1,i),(2,j)}(\bm{x}_1, \bm{x}_2 ;t) \mathrm{Enc}(\mathrm{encounter};\bm{x}_1, \bm{x}_2,t) \\
&+ \sum_{i=1}^{N_{X1}} \sum_{j=1}^{N_{X2}} \sum_n \sum_{k=1}^{N_{Xn}} k_{\mathrm{evt}} \int d^3 \bm{x}_1 d^3 \bm{x}_2 d^3 \bm{x}_3 \; f_{(1,i),(2,j),(n,k)}(\bm{x}_1, \bm{x}_2, \bm{x}_3 ;t) \mathrm{Enc}(\mathrm{encounter};\bm{x}_1, \bm{x}_2,t) + \dots
\end{aligned}
\end{equation}

\subsection{Significance of weight-experimental value comparison}
Application of the canonical distribution framework to equation (\ref{delta_s}) yields the binding probability:
\begin{equation}
\begin{aligned}
p &= \frac{\exp \left(-\frac{\Delta F}{k_B T}\right)}{1+\exp \left(-\frac{\Delta F}{k_B T}\right)} \\
&= \frac{\exp \left(-\frac{\Delta E}{k_B T}+\frac{\Delta S}{k_B}\right)}{1+\exp \left(-\frac{\Delta E}{k_B T}+\frac{\Delta S}{k_B}\right)} \\
&= \frac{\exp \left(-\frac{\Delta E}{k_B T} + \frac{\Delta S_{\mathrm{internal}}}{k_B} + \frac{\Delta S_{\mathrm{inh}}}{k_B} \right) w(X_1, X_2) }{1+\exp \left(-\frac{\Delta E}{k_B T} + \frac{\Delta S_{\mathrm{internal}}}{k_B} + \frac{\Delta S_{\mathrm{inh}}}{k_B} \right) w(X_1, X_2)} .\\
\end{aligned}
\end{equation}
In the weak-binding regime where $\exp \left(-\frac{\Delta F}{k_B T}\right) \ll 1$, the binding probability $p$ approximates to:
\begin{equation}
\begin{aligned}
p \simeq \exp \left(-\frac{\Delta E}{k_B T} + \frac{\Delta S_{\mathrm{internal}}}{k_B} + \frac{\Delta S_{\mathrm{inh}}}{k_B} \right) w(X_1, X_2).
\end{aligned}
\end{equation}
Thus, the binding probability $p$ and weight $w(X_1,X_2)$ exhibit a proportional relationship.

While both $w(X_1,X_2)$ and $\Delta S_{\mathrm{inh}}$ depend on the probability distributions $f_{X1}(x)$ and $f_{X2}(x)$, $\Delta S_{\mathrm{inh}}$ is invariant under renormalized linear changes of variables that preserve normalization,
\begin{equation}
f(\bm{x}) \rightarrow |\det A|\; f(A\bm{x}),
\end{equation}
where $A$ is an invertible linear transformation and $|\det A|$ ensures $\int f(\bm{x})\,d\bm{x}=1$.

In the present simulations, parameter variations altered only distribution widths while preserving shapes, maintaining $\exp (\Delta S_{\mathrm{inh}})$ as approximately constant. Therefore, the proportional relationship between binding probability $p$ and weight $w(X_1,X_2)$ persists even after $\exp (\Delta S_{\mathrm{inh}})$ corrections.

Validation of the weight-based approach requires experimental systems operating within the regime $\exp \left(-\frac{\Delta F}{k_B T}\right) \ll 1$, enabling direct assessment of proportionality with experimental values.
Weights fundamentally quantify encounter frequencies, and their correlation with experimental binding is non-trivial; nonetheless, the observed proportionality supports the theoretical framework.
Selection of systems with moderately weak binding was important: strong binding in confined geometries would lead to complete saturation, eliminating the discriminatory power necessary for meaningful comparison.

\putbib[ref]
\end{bibunit}

\end{appendices}

\end{document}